\documentstyle[12pt,aasms4,psfig]{article}

\def\,{\thinspace}
\def \kms{km\,s$^{-1}$}

\newcommand{\solm}{M$_{\odot}$}

\newcommand{\CO}{$^{12}$CO }
\newcommand{\COe}{$^{12}$CO (1--0) }
\newcommand{\COz}{$^{12}$CO (2--1) }

\newcommand{\rf}{\par\noindent\hangindent 15pt {}}
\newcommand{\apjj}[2]{ApJ, #1, #2.}
\newcommand{\apjjs}[2]{ApJS, #1, #2.}
\newcommand{\apjjl}[2]{ApJ, #1, #2.}
\newcommand{\asa}[2]{A\&A, #1, #2.}
\newcommand{\asal}[2]{A\&A, #1, #2.}
\newcommand{\asas}[2]{A\&AS, #1, #2.}

\newcommand{\ppasp}[2]{PASP, #1, #2.}
\newcommand{\appss}[2]{Ap\&SS, #1, #2.}
\newcommand{\mn}[2]{MNRAS, #1, #2.}
\newcommand{\ajj}[2]{AJ, #1, #2.}

\newcommand{\vol}[1]{1}
\newcommand{\nhico}{$\frac{N_{H_2}}{I_{CO}}$}

\lefthead{Schinnerer et al.}
\righthead{Molecular Gas in NGC~1068}

\begin{document}

\title{Bars and Warps traced by the Molecular Gas in the 
Seyfert~2 Galaxy NGC~1068}

\author{E. Schinnerer \altaffilmark{1}, A. Eckart, L.J. Tacconi, R. Genzel}
\affil{Max-Planck-Institut f\"ur extraterrestrische Physik, 
85740 Garching, Germany}
\and 
\author{D. Downes}
\affil{Institut de Radio Astronomie Millim\'etrique, 38406 Saint Martin
d'H\`eres, France}

\altaffiltext{1}{Present address at: California Institute of Technology, Pasadena, 
CA 91125}

\altaffiltext{2}{Ap.J. accepted}

\begin{abstract}
We present new interferometer observations of the \COe and \COz line
emission of NGC~1068 with a resolution of $0.7''$. 
The molecular gas in the inner
$5''$ is resolved into a ring with two bright knots east and west of
the nuclear continuum emission. 
For the first time
in NGC~1068, we can trace molecular gas at $\approx 0.18''$ (13~pc) from
the nucleus. The high velocities in this region imply an enclosed mass
of $\sim$10$^8$ \solm .  This value is consistent with a black hole
mass of $1.7\times 10^7$\,\solm , as estimated from nuclear H$_2$O
maser emission, plus a contribution from a compact nuclear stellar
cluster.
Perpendicular to the kinematic major axis optical images of NGC~1068 
show a bright, stellar, oval structure of eccentricity 0.8 and a deprojected 
length of 17~kpc.
Analysis of the rotation curve shows the CO spiral arms are at the 
inner Lindblad resonance of this bar-like structure.  
Inside the molecular spiral arms, $10''$ from the nucleus,
the CO kinematic axis changes direction  probably in response to the
to the 2.5~kpc (deprojected) long stellar bar seen in the near infrared (NIR).
The low velocity dispersion indicates the
molecular gas is in a disk with a thickness of 10~pc in the nuclear
region and 100~pc in the spiral arms.

We constructed kinematic models for the molecular gas using elliptical orbits
caused by a $\sim 1''$ (72~pc) nuclear bar and using tilted rings
resulting in a warp.  We find that the gas motions are consistent 
with either the warp or the bar models.
However, because there is no evidence for a $\sim 1''$ nuclear bar in
NIR images, we favor the warp model. A warped CO disk can also explain
the obscuration of the AGN, the extinction of light from the nuclear
stellar cluster, and the observed NIR and mid-IR polarization. The
model predicts the warped CO disk should become edge-on at a radius of
70~pc, thereby creating a cavity for the ionization cone.

\end{abstract}


\section{INTRODUCTION}
Knowledge of the distribution and kinematics of the circumnuclear
molecular gas in Seyfert galaxies is essential for understanding the
fueling of the central region and the role of gas and dust in
obscuring the active galactic nucleus (AGN).  Recent improvements in
millimeter interferometry now allow us to observe molecular lines with
sub-arcsecond resolution, high spectral resolution, and high
sensitivity.  In nearby Seyfert galaxies, with the best resolution
currently available, we can now search for molecular gas down to radii
of a few tens of parsecs.  Models of the motions of this molecular gas
can then help us to understand the mechanism that transports gas
into the center.

An overview of the modeling results so far is given in Schinnerer, Eckart \&
Tacconi (1999a; hereafter SET99a), and a detailed model of the nuclear
molecular gas in the Seyfert~1 galaxy NGC~3227 is given by Schinnerer,
Eckart \& Tacconi (1999b; hereafter SET99b). In the inner 80~pc of
NGC~3227, for example, the kinematics of the molecular gas are better
explained by a warping of the thin gas disk than by motion in the
potential of a nuclear stellar bar. In that galaxy, molecular gas is
detected down to a radius of $\sim$ 10\,pc.

In this paper, we model the complex kinematics of the central few
hundred parsecs of the Seyfert~2 galaxy NGC~1068.  This galaxy is at a
distance of 14.4\,Mpc ($H_0$ = 75\,\kms ; see Bland-Hawthorn et
al. 1997), giving a scale of 72\,pc arcsec$^{-1}$ ({\bf Table
\ref{vv40}}), which is very favorable for detailed studies of the
circumnuclear gas.  NGC~1068 was originally classed as a Seyfert~2
galaxy based on its direct-path, i.e. unpolarized emission lines, which
have narrow widths (Khachikian \& Weedman 1974). This narrow-line
region (NLR) is extended to the northeast of the nucleus and coincides
with the radio jet.  Antonucci \& Miller (1985) later showed that the
scattered-path, polarized emission lines have widths typical of
broad-line regions (BLR), indicating that NGC~1068 has a hidden
Seyfert~1 nucleus.  Around this nucleus, there is a central nuclear
star cluster with a FWHM diameter of $0.6''$ (43\,pc) that was mapped
by Thatte et al. (1997).  On a much larger scale, there is a stellar
bar 2.3\,kpc long, detected in the NIR (Scoville et al. 1988;
Thronson et al. 1989).  This bar is in turn surrounded by a
circumnuclear starburst ring that, along with other star forming
regions, contributes half the bolometric luminosity of the galaxy
(Telesco et al. 1984; Davies et al. 1998).  This starburst ring is
also detected in lines of molecular gas (Planesas et al. 1991; Jackson
et al. 1993; Tacconi et al. 1994; Helfer \& Blitz 1995; Tacconi et
al. 1997) and in the radio continuum (e.g. Gallimore et al. 1996).

The plan of this paper is as follows.  Section 2 describes the
observations and data reduction.  Section 3 covers the distribution
and kinematics of the CO emission, the millimeter continuum, and the
location of the nucleus.  In section 4 we derive the rotation curve of
the molecular gas and discuss its relation to the stellar bar and the
location of dynamical resonances.  We then derive the molecular gas
mass and the total dynamical mass (section 5) and derive the thickness
of the molecular gas disk from the velocity dispersion (section 6).
In section 7, we fit the kinematic data with bar and warp models.  and
discuss the model results in section 8.  Section 9 summarizes our
conclusions.

\section{
\label{int}
OBSERVATIONS AND DATA REDUCTION}
The \CO lines were observed in February/March 1997 with the IRAM
interferometer on Plateau de Bure, France, with 5 antennas in the
long-baseline A, B1, and B2 configurations and in November 1995 to
March 1996 with 4 antennas in the shorter-baseline C1, C2, and D1
configurations. The \COe observations had baselines from 24 to 408\,m
that gave a beam of $2.2'' \times 1.2''$ at p.a.~26$^\circ$ after
uniform weighting. The \COz observations had only the A, B1, and B2
baselines, ranging from 40 to 408\,m, that gave a beam of $1.0''
\times 0.5''$ (p.a.\ 26$^\circ$) after uniform weighting.  The
bandpass was calibrated on 3C454.3 and 3C345.  The source 0235+164 was
observed every 10\,min for phase and amplitude calibration. The data
were CLEANed with 700 iterations and convolved to beams of $1.4''
\times 1.4''$ at \COe and $0.7'' \times 0.7''$ at \COz, with velocity
channels of 10\,\kms\ for both lines.  We also made a 115\,GHz
continuum map from the line-free channels from $-$310 to
$-$190\,\kms\ and +200 to +320\,\kms . To estimate the line flux, we
subtracted the mean continuum from the channels with (line+continuum).

Our \COe map agrees well with CO maps made with
other interferometers, and our integrated \COe flux agrees with that
in the interferometer spectrum of Helfer \& Blitz (1995) to within
15\%.  Helfer \& Blitz (1995) also combined their BIMA array and NRAO
single dish data. Their BIMA/NRAO map has 25 to 30\% more flux than
their BIMA map alone, so some of the large-scale flux may be resolved
out by the BIMA and IRAM arrays.  To check our calibration of the \COz
line emission, we made a \COz map with the same beam as our \COe
map. At a resolution of $1.4''$, the $\frac{{\rm CO}(2-1)}{{\rm
CO}(1-0)}$ ratio for the nuclear ring is 1.0 ($\pm 20$\%), in
agreement with the ratio of 1.1 found in this region with the IRAM
30\,m telescope (Braine \& Combes 1992).

\section{
\label{ecc10}
MOLECULAR GAS DISTRIBUTION AND PROPERTIES}
\subsection{The distribution of the line emission
}
\begin{figure}
\begin{center}
\psfig{file=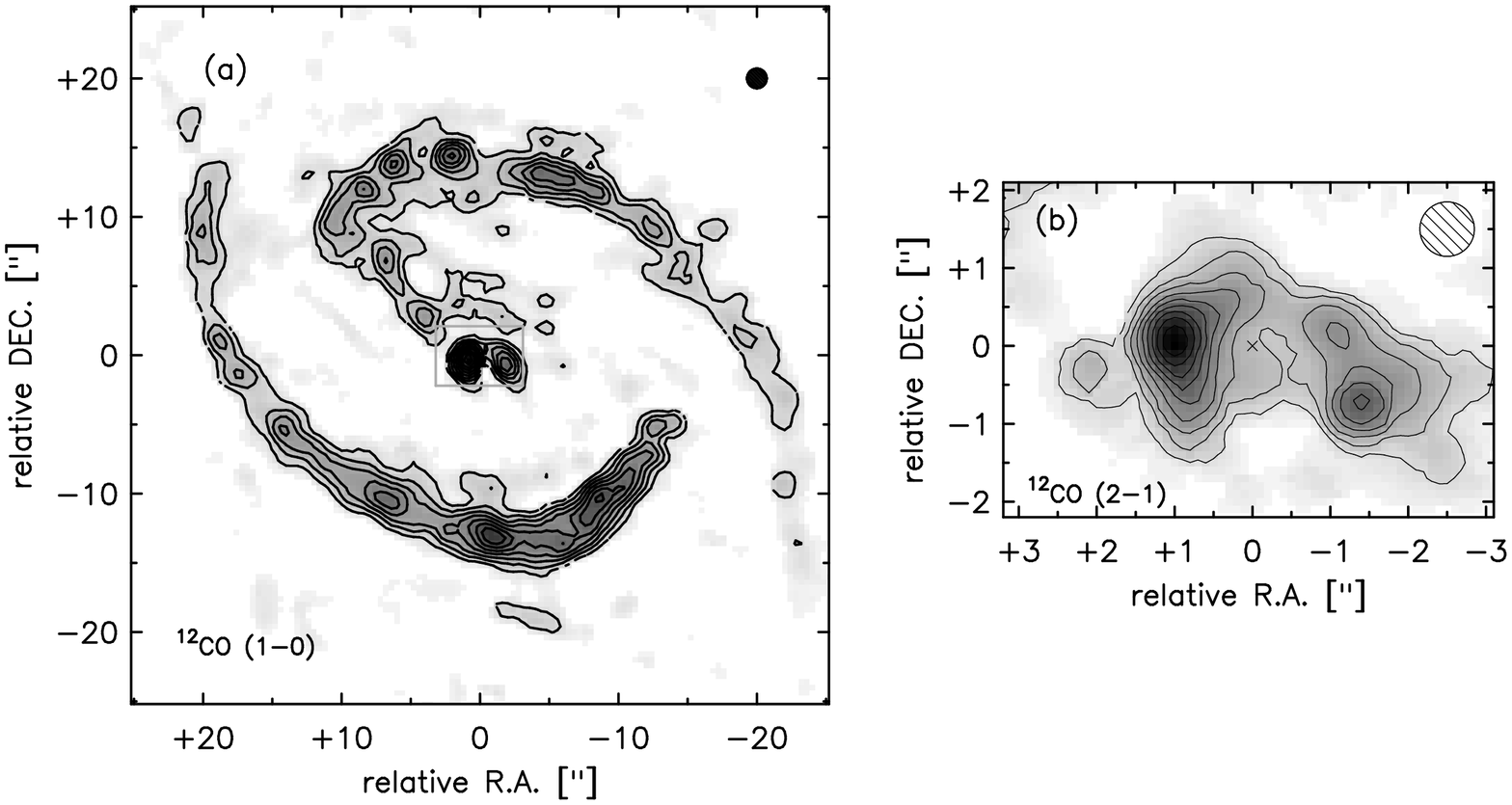,height=10.0cm,width=14.1cm,angle=-90.0}
\end{center}
\figcaption[]{
\label{vv1}
Maps of {\it (a)} \COe  and {\it (b)} \COz  in the center of NGC~1068. 
Contours are at 10, 20, ... 100\% of the peak, which is 1.24 Jy km/s/beam
in \COe and 3.43 Jy km/s/beam in \COz. The central box in the \COe map 
shows the area of the \COz map.  
}
\end{figure}

Our new observations allow us to study the millimeter molecular line
and continuum emission at an unprecedented angular resolution.
The \COz data, in particular, give very detailed maps of the molecular gas
in the inner 200~pc of NGC~1068.
At 100~pc from the center, the \COz emission is in a ring (hereafter
the {\it nuclear ring}; {\bf Fig. \ref{vv1}}). In this nuclear ring, there
are two emission knots, one at $1''$ east of the nucleus and another
one $1.5''$ west and $1''$ south of the center. At $\pm 0.5''$ north
and south of the center, there are emission ridges connecting the two
knots at 25\% of the peak line intensity. This same overall nuclear
structure also exists in the CS(2--1) line (Tacconi et al. 1997) and
the HCN(1--0) line (Tacconi et al. 1994).  The $\frac{{\rm CO}
(2-1)}{{\rm CO}(1-0)}$ ratio is between 0.8 and 1.0 over the entire
nuclear ring, except for a part of the western emission knot, which
has a ratio of 2.0.  This is also evident in the nuclear spectra. The
\COz profile has three peaks while the \COe profile has only two. The
missing peak in the \COe line is offset by +75\,\kms\ relative to the
systemic velocity and is associated with the western knot. This
suggests the CO lines are optically thin in parts of the western knot.

As shown in earlier maps, most of the \COe emission is in two spiral
arms with a diameter of $\sim 40''$ ({\bf Fig. \ref{vv1}}; Planesas et al.
1991; Kaneko et al. 1992; Helfer \& Blitz 1995). On our new map, 
the \COe peak $\sim 1''$ east of the center is 1.5 times stronger than
the spiral-arm emission. In our map, the northern spiral arm is less
luminous than the southern one, in contrast to the earlier
measurements of Helfer \& Blitz (1995; see below).

CO emission is also observed along the NIR stellar bar found by
Scoville et al. (1988) at about half of the nuclear peak
intensity. With our $1.4''$ beam, we mainly find emission in the
northern half of the bar. In the lower resolution ($\sim 3.4''$) \COe
data (Tacconi et al. 1997) emission is seen on both sides of the bar,
but stronger north of the nucleus. Both sides of the bar were also
detected in lower resolution \COe maps by Helfer \& Blitz (1995) and
in the $^{13}$CO(1--0) map by Tacconi et al.  (1997). The tips of the
NIR stellar bar almost touch the spiral arms. The spiral arms are
not detected in the \COz map, because their diameter is larger than
the $22''$ primary beam of the IRAM antennas, and a mosaic map would
be needed to recover the spiral arms in \COz.  Furthermore, the spiral
arms are weaker than the nucleus and they may be partly resolved since
we only used long baseline configurations for the \COz mapping.

Our \COe intensity map agrees well with the OVRO map (Baker \& Scoville
1998; private communication).  Differences with the earlier data of
Helfer \& Blitz (1995) may be due to uncertainties in the calibration
and deconvolution of the BIMA/NRAO data and the possibility that 30\%
of the extended line flux is resolved out in our interferometer map.

\subsection{Radio continuum emission}
\begin{figure}
\begin{center}
\psfig{file=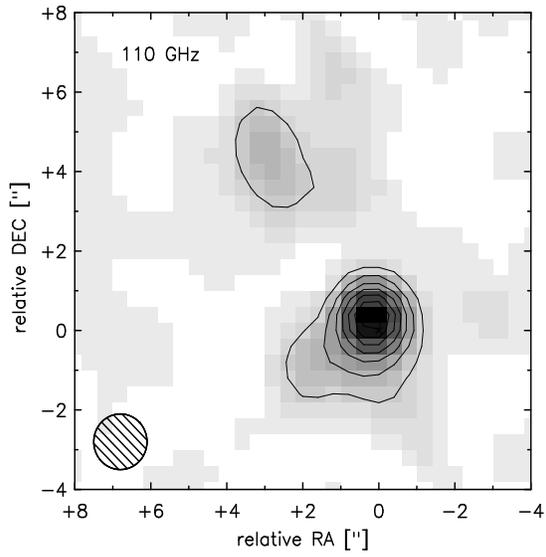,height=14.1cm,width=10.0cm,angle=0.0}
\end{center}
\figcaption[]{
\label{vv2}
The distribution of the 110\,GHz continuum in NGC~1068 with a
sensitivity of 1$\sigma$ = 1 mJy/beam. The contours
start at 3$\sigma$ in steps of 3$\sigma$. The peak intensity is 23 mJy/beam.
}
\end{figure}

Our 115\,GHz continuum map shows two emission knots, one coinciding
with the nucleus, and the other $2.5''$ east and $4.5''$ north of the
nucleus ({\bf Fig. \ref{vv2}}). The nuclear knot is slightly resolved
by our $1.4''$ beam. A Gaussian fit gives $1.5'' \times 1.7''$ at
p.a.\ 156$^\circ$. At 115\,GHz, the nuclear knot has a flux density of
(36$\pm$5)\,mJy, while the northern knot has (15$\pm$5)\,mJy.  The
northern knot coincides with the peak of the 5\,GHz radio jet mapped
by Wilson \& Ulvestad (1983), as noted previously by Helfer \& Blitz
(1995), who observed the same two continuum knots in their BIMA map at
85\,GHz.  This structure at 85\,GHz is also observed with the IRAM
array (Tacconi 1999; private communication).  At 85\,GHz, both knots
have the same flux density, 40\,mJy.  At 230~GHz, no continuum
emission was detected, to a limit of 6\,mJy at each knot.

\section{GLOBAL KINEMATICS OF THE MOLECULAR GAS}
In this section, we describe the velocity field of the molecular gas,
estimate the enclosed nuclear mass, and derive a rotation curve.  We
then combine this rotation curve with optical and IR evidence for bars
in NGC~1068 to deduce the locations of dynamical resonances.

\subsection{
\label{ecc2}
The large-scale CO velocity field
}
\begin{figure}
\begin{center}
\psfig{file=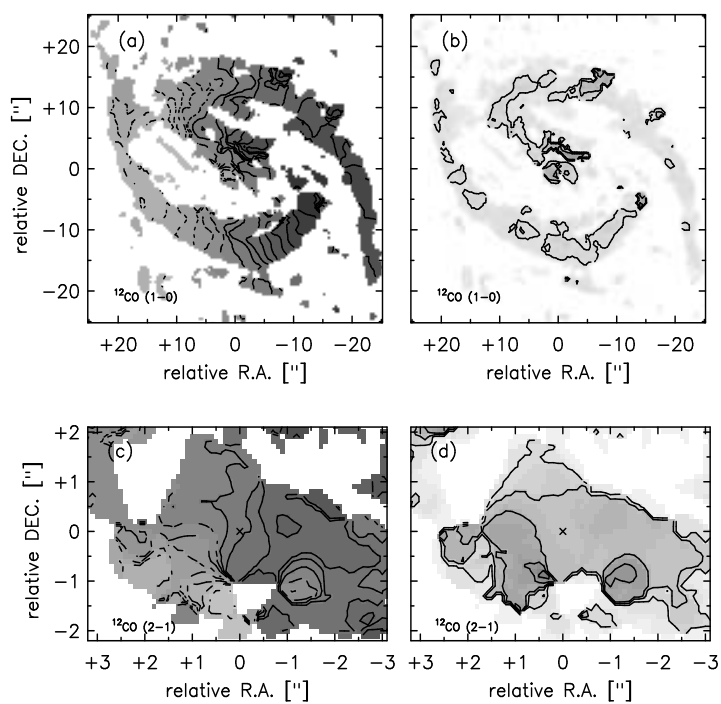,height=14.1cm,width=10.0cm,angle=0.0}
\end{center}
\figcaption[]{
\label{vv3}
Velocity field and velocity dispersion of the \CO lines in NGC~1068.
The velocity field (a) and the velocity dispersion map (b) of the \COe
emission (beam $1.4''$) and the \COz emission (beam $0.7''$)
(c and d) are shown, respectively. The contours of the velocity
fields are in steps of 20\,\kms\ (v=0\,\kms\ corresponds to the first
central solid line next to a broken line). In the dispersion maps, 
contours are at 15, 30, 45 and
60\,\kms\ for \COe  and at 15, 30, 45, 60 and 75\,\kms\ for the
\COz.
}
\end{figure}

A striking feature of the \COe velocity field is the change of
position angle of the kinematic major axis within the spiral arms
({\bf Fig.\,\ref{vv3} a and c}).  At radii $> 15''$ the velocity field
of the molecular gas joins the outer HI velocity field at
p.a. $280^\circ$ and inclination $40^\circ$ (Brinks et al. 1997).  At
$r < 10''$, the kinematic major axis changes its position angle from
278$^\circ$ to 305$^\circ$. The change in position angle at $r \sim
10''$ also occurs in the velocity fields of the H$\alpha$ (Dehnen et
al. 1997) and the stars (Garcia-Lorenzo et al. 1997). This effect also
appears in the HCN(1--0) line (Jackson et al.  1993; Tacconi et
al. 1994) and was analyzed by Helfer \& Blitz (1995) in \COe.

\subsection{
\label{ecc1}
The nuclear kinematics from the CO position-velocity diagrams
} 
\begin{figure}
\begin{center}
\psfig{file=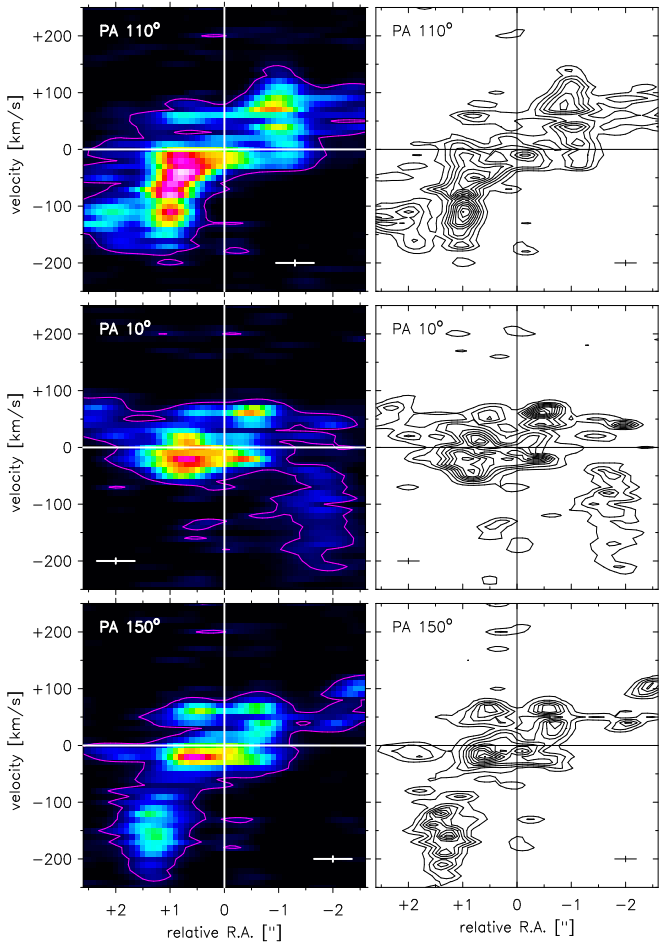,height=14.1cm,width=10.0cm,angle=0.0}
\end{center}

\figcaption[]{
\label{vv4}
Position-velocity diagrams of the nuclear \COz line emission in NGC~1068
along different position angles. 
To highlight this complex velocity behavior the data is shown at a nominal 
resolution of 0.4'' (right).
The contours are at 10, 30, ... 100\% of the peak.  
For comparision the corresponding pv-diagrams at the achieved 
instrumental resolution of 0.7'' are shown in color on the left panels.
The velocity resolution in both cases is 10 \,\kms.
The (magenta) contour corresponds to 3$\sigma$, with 1$\sigma$= 6.0
mJy/beam.
A remarkable feature in the
$pv$-diagram along p.a.\ 110$^\circ$ is the large velocity dispersion
at $\pm 1''$ from the center. 
For the first time
in NGC~1068, we can trace molecular gas at $\approx 0.18''$ (13~pc) from
the nucleus. The high velocities in this region imply an enclosed mass
of $\sim$10$^8$ \solm ~not correcting for inclination effects.
This value is consistent with a black hole
mass of $1.7\times 10^7$\,\solm , as estimated from nuclear H$_2$O
maser emission  (Greenhill \& Gwinn 1997), plus a contribution from
a compact nuclear stellar cluster (Thatte et al. 1997).
}
\end{figure}

In the inner 500\,pc, the \COz velocity field is perturbed by the two
bright knots $\sim 1''$ east and west of the nucleus that do not
follow the overall circular rotation.  To more easily separate
symmetric and asymmetric components and see which bits of the
molecular gas have ordered motions, we used position-velocity diagrams
($pv$ diagrams) along different angles.  We made $pv$ diagrams every
20$^\circ$ starting at 10$^\circ$ ({\bf Fig. \ref{vv4}}).  The diagram
along p.a.\ 110$^\circ$ is remarkably symmetric and has most of the
flux, because it crosses both emission knots.  All diagrams close to
the kinematic {\it minor} axis are compact, because there is only weak
emission north and south of the center.

In \COz, the western knot (knot W) is slightly farther from the
nucleus ($\sim 1.5''$) than the eastern knot.  Knot W appears in $pv$
diagrams at p.a.\ 50$^\circ$ and 70$^\circ$, with a velocity
dispersion of 60\,\kms .  Knot E, with a velocity dispersion of
70\,\kms , appears in $pv$ diagrams at p.a.  150$^\circ$, 170$^\circ$,
0$^\circ$, and 10$^\circ$ as a distinct component from $-$230 to
$-$30 \,\kms\ at $\sim 1.5''$ southeast of the center.  To estimate
these components' flux, we calculated the moments only for velocities
$-$230 to $-$30\,\kms . The apparent dispersion of knot W is a mixture
of the 40\,\kms\ velocity spread of the nuclear disk with the
30\,\kms\ spread of the knot itself.  For the eastern knot, it is hard
to see the knot component, since it is blended with the larger-scale
rotating disk.  These two knots ({\bf Table \ref{vv42}}) contain a few
10$^6$\,\solm\ of molecular gas, which suggests they are giant
molecular cloud complexes.  There is a similar mixture in the
circumnuclear molecular gas in NGC~3227, where we also found several
components deviating from the overall circular rotation (SET99b).

\subsection{
\label{cc1}
Hints of a nuclear enclosed mass}
All position-velocity diagrams show significant high velocity emission
at $r \leq 0.2''$, similar to that in the nucleus of NGC~3227
(SET99b).  If the high velocities trace rotation in the immediate
vicinity of the nucleus, we can estimate the enclosed nuclear mass.
The minimum radius at which we detect an emission peak of 
high-velocity nuclear molecular
gas is 13.5\,pc, and its mean velocity relative to systemic is $\Delta
v \approx 188$\,\kms , uncorrected for inclination. If this is
rotation, it implies a mass $\geq 1.5 \times 10^8$\,\solm\ within the
inner 27\,pc diameter, comparable with the mass of $4.5 \times 10^8$\,\solm\
within the inner 70\,pc of our Galaxy (Genzel et al. 1994).  This is
ten times the black hole mass of $1.5 \times 10^7$\,\solm\ estimated
from H$_2$O masers in the inner 2\,pc of NGC~1068 (Greenhill \& Gwinn
1997).
The mass in the inner 27\,pc is thus mostly due to a central,
compact part of the $0.6''$ (43\,pc) stellar cluster as mapped
by Thatte et al. (1997).

\subsection{
\label{ecc4}
The rotation curve
}
The molecular gas velocity field inside the spiral arms is complex,
making it hard to derive the rotation curve from \CO alone.  We now
discuss our best estimate of the CO rotation curve and compare our
results with the stellar motions.  We show that the velocity fields of
gas and stars are similar, reflecting the influence of the bar on the
motions at $3''< r < 15''$.  We think the difference in the rotation
curves of the gas and stars could be due to the complex dynamics
induced by the stellar bulge.

\subsubsection{\label{ecc5}Rotation curve from gas motions
}
\begin{figure}
\begin{center}
\psfig{file=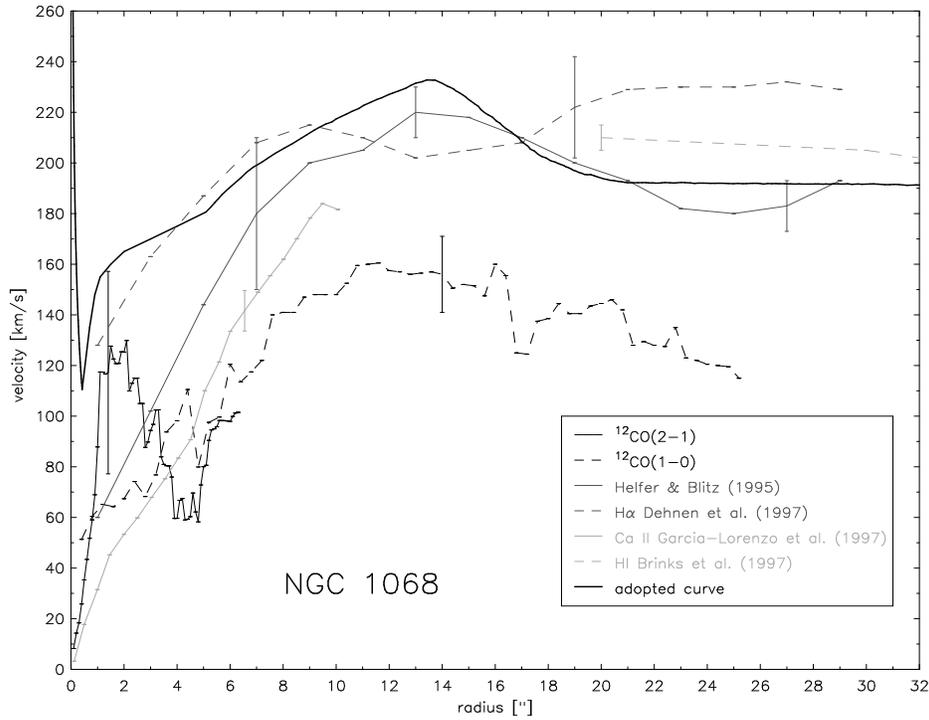,height=10.0cm,width=14.1cm,angle=-90.0}
\end{center}
\figcaption[]{
\label{vv5}
Rotation curves for NGC~1068. 
The \CO curves are the average values 
of the minimum and maximum velocity at any given radius 
(uncorrected for inclination). The other curves are taken from the
quoted references. The increase in velocity of the \COz curve 
at a radius of $r \sim 1''$ can also be seen in the $pv$ diagrams of the
observed data. At radii between $10''$ and $16''$ the rotation curve 
agrees with fits of a rotating gas disk to the observed data. 
}
\end{figure}

The \CO velocity field in the center of NGC~1068 deviates strongly
from circular rotation at $r < 13''$, so a rotation curve derived from
a rotating disk model would probably be wrong.  The velocity field and
$pv$ diagrams show however, that most of the \CO emission is indeed
from a centro-symmetric structure. We therefore used the first-moment
velocity maps of the \COe and \COz emission to determine the velocity
extrema in circular annuli around the center. We chose the \COz
velocity map because the high-velocity-dispersion knots on either side
of the nucleus are partly resolved ({\bf Figs.\,\ref{vv1} and
\ref{vv4}}).  The mean velocities in these maps are lower limits to
the true rotation speeds since we did not correct for inclination. Our
mean curve, corrected for an inclination $i = 40^\circ$, agrees well
with published rotation curves within the uncertainties ({\bf Fig.\
\ref{vv5}}).  We compared our rotation curve with that derived from
H$\alpha$ (Dehnen et al. 1997; errors 10 to 30\,\kms\ for $r < 16''$)
and with the curve derived from the BIMA \COe
data (Helfer \& Blitz 1995; errors 30 to 70\,\kms\ for $r < 16''$ and
10 to 20\,\kms\ for $16'' \leq r < 30''$).
 
From $r=2.3''$ moving inward, our \COz rotation curve first increases
by 50\,\kms\ and then decreases again at $r = 1''$.  This range of
radii contains the nuclear ring and the two barely-resolved CO knots
east and west of the nucleus.  The $pv$ diagrams along p.a.\
110$^\circ$ are similar in \COe and \COz.  We derived the rotation
curve at $r < 1''$ for two cases: (1) rotation velocities go to
0\,\kms\ at the center, and (2) Keplerian velocities around a point
mass interior to $r = 0.4''$ (section \ref{cc1}).

\subsubsection{The stellar velocity field
}
Garcia-Lorenzo et al. (1997) mapped the stellar velocities in the
inner $24'' \times 20''$ with the optical Ca II triplet at $\lambda$
8498, 8575, and 8695\AA. The velocity field of the stars matches that
of the \COe emission to within 50\,\kms\ and $1''$.  This indicates
that at $r > 3''$ (the resolution of the Garcia-Lorenzo et al. data),
the stars and gas have similar kinematics.  The velocity dispersion
observed in the Ca~II lines agrees well with the velocity dispersion
derived in the NIR (Dressler 1984; Terlevich et al. 1990; Oliva et
al. 1995), suggesting that the velocity field measured in the Ca~II
line is indeed representative of the overall nuclear region (inner 15'' to 
20'') and does not
merely trace the surface of an otherwise obscured stellar system.

For both stars and gas, at radii $\leq 10''$, the isovelocity lines in
the north bend to the east, and the isovelocity lines in the south
bend to the west.  Such a bending is as expected from our model
calculations (section \ref{ecc6}) for orbits induced by a bar.  It
also agrees with the expected behavior of gas and stars in the
potential of the NIR bar (e.g., Athanassoula 1992; Maciejewski \&
Sparke 1999). To confirm this finding observationally, more data are
needed on the Ca~II velocity field south ($\sim 8''$) of the nucleus,
since the bending in the isovelocity lines occurs at a higher velocity
relative to systemic than in the north.  Higher-sensitivity molecular
line data south of the nucleus would also be useful.

\subsubsection{
\label{eec0}
Rotation curve in the inner $6''$}
In the inner region the kinematics of stars and gas may be decoupled.
The stellar kinematic data (Garcia-Lorenzo et al. 1997) 
indicate that the stellar rotation curve lies well below the
rotation curve derived from the \CO data.
This probably reflects the fact that in the inner $6''$
the velocity dispersion of the stars is
higher relative to that of the molecular gas.
In this region, the gravitational force of the gas is negligible relative to
that of the stars, so we estimate the effects of different stellar
potentials as follows.  (1) We start with the simplest model, an
isotropic, spherical stellar cluster. (2) We then generalize our model
to central oblate stellar systems. (3) Finally, we consider the effect
of a stellar bar on stellar orbits in the center of NGC~1068.

{\it Isotropic, spherical stellar cluster:}
The mass of an isotropic, spherical, non-rotating stellar cluster can be
estimated from the stellar velocity dispersion and the Jeans equation,
derived from the collisionless Boltzmann equation (see Binney \& Tremaine 1987). 
Thatte et al. (1997) assumed a
velocity dispersion of 140\,\kms\ (Dressler 1984) to estimate a mass
of $6.5 \times 10^8$\,\solm\ within the inner $2''$ diameter.  This is
nearly twice our value of $3.7\times 10^8$ \solm\ for the same region
derived from the molecular gas. The discrepancy could be due to 
(1) the stellar flux coming from supergiants with deeper Ca II absorption
than in the template giant stars, yielding a lower apparent velocity
dispersion, or 
(2) the velocity dispersion of the underlying bulge component, which
emits up to 45\% of the light in the inner $5''$ (Table 1 of Thatte et
al. 1997). Comparison of $I$-band data in the inner $15''$, which show
constant velocity dispersion (Dressler 1984), with the nuclear $H$ and
$K$ band data indicates that the velocity spread of the nuclear
cluster cannot exceed that of the outer zone. All this suggests the
mass derived by Thatte et al. (1997) is an upper limit to the enclosed
nuclear mass.

{\it Central oblate system:}
Observations and theory suggest bars turn into bulges at small radii
(Norman et al. 1996; Hasan et al. 1993; Combes et al. 1990). 
For these systems the Jeans equation separates into two
equations giving the vertical and radial form of the gravitational
potential (see review by Merritt 1999).
Binney \& Tremaine (1987, their Fig.\ 4-5) show that 
if self-similar isodensity ellipsoids of stellar spheroidal systems 
become more spherical at smaller radii then $v/\sigma$ falls below unity.
This is indeed indicated by $K$-band images for $r \leq 5''$
(e.g., Fig.\ 1 in Thatte et al. 1997).
Hence if the observed stellar velocity dispersion is constant, the circular
velocity must decrease at smaller radii. This is consistent with the
observed drop in the stellar rotation curve toward the center of
NGC~1068.

{\it Influence of a bar potential:}
NGC~1068 has a stellar bar in the inner $20''$ seen in the NIR
(Scoville et al. 1988; Thronson et al. 1989). Bars have prolate
symmetry and are probably triaxial. Such a system can have an
anisotropic velocity dispersion and its ellipsoid need not be coaxial
with the bar (Fillmore 1986). It is thus not easy to predict the
velocity dispersion, since in a bar potential, many orbit families are
extended in the $z$-direction and there can be radial and vertical
resonances that support specific orbit families, affecting the stellar
velocity dispersion (e.g. Pfenniger 1984; Combes et al. 1990;
Olle \& Pfenniger 1998). 
Unlike the stars in the center of NGC~1068, the molecular gas has a
low velocity dispersion (section \ref{ecc2}), indicating it is
decoupled from the stars. This explains why the gas rotation curve ---
which we used for our models --- lies above the stellar rotation curve
by Garcia-Lorenzo et al. (1997) at small radii. Hence, for circular
orbits, the molecular gas is a good tracer of the potential and its
rotation curve is governed by the enclosed dynamical mass.

\subsection{
\label{ecc8}
The bars and resonances}
The interaction of the molecular gas with the stellar gravitational
potential leads to  resonances that in turn determine the
distribution and kinematics of the gas. We now discuss the evidence
for bars and the locations of dynamical resonances in NGC~1068.

\subsubsection{The Bars}
Although NGC~1068 is not normally considered a barred spiral, Scoville
et al. (1988) nevertheless detected at $K$ band a $32''$ long bar at
p.a.\ $\sim 48^\circ$.  The ends of the NIR stellar bar join the
spiral arms in the molecular gas (e.g., Tacconi et al. 1997). In most
bar theories, there are no massive pile-ups of gas at the corotation
radius near the ends of the bar. These accumulations are only
expected at the inner and outer Lindblad resonances. The fact that the
arms are at the bar corotation radius is thus puzzling and could indicate 
the dynamical effect of a second, larger bar in NGC~1068.
(Here, we assume that the corotation lies about at the end of the bar as
found for late type galaxies, see Combes 1997 for a review.)

\begin{figure}
\begin{center}
\psfig{file=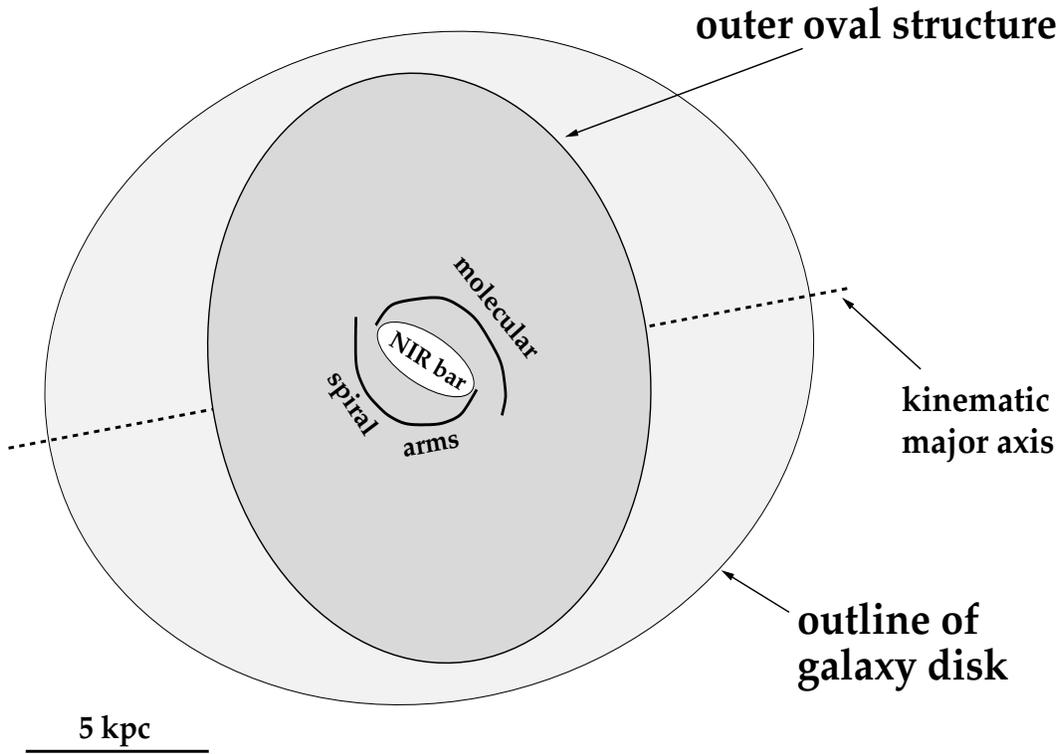,height=10.0cm,width=14.1cm,angle=-90.0}
\end{center}
\figcaption[]{
\label{abb99}
Sketch of NGC~1068 showing the location of the bars and molecular
spiral arms. With the exception of the outline of the galaxy disk the 
figure is approximately to scale.
}
\end{figure}

Optical images of NGC~1068 show a bright inner disk at radii $<
100''$, with spiral arms and dust lanes extended north-south (see
sketch in Fig. \ref{abb99}). This
oval structure is $180''$ ($\sim 13$\,kpc) long, at p.a.\ 5$^\circ$,
with an eccentricity of 0.8 (Bland-Hawthorn et al. 1997, from $BRI$
images by R.B.\ Tully).  At radii $> 15''$, however, the {\it
kinematic} major axis runs almost exactly east-west. Hence the
eccentricity of the oval structure on the sky is not due to the
galaxy's inclination; instead, this structure is {\it a large-scale
bar along the minor axis of NGC~1068}. In the Palomar Sky Survey
images of NGC~1068, the outer, weak emission (radii $> 100''$) is
extended east-west, and the eccentricity of its lowest contours
corresponds to the 40$^\circ$ inclination derived from HI velocities
(Brinks et al. 1997). Deprojection of the optical image then yields a
bar length of $240''$ (17.3\,kpc). As shown in section \ref{ecc20},
this bar has its inner Lindblad resonance (ILR) at $r= 18''$
(1.3\,kpc), near the molecular spiral arms. The outer bar-like
structure in NGC~1068 could thus be the cause of these spiral arms.
Further support for this idea comes from the deprojected shape of the
molecular spiral arms, which is an almost perfect ring --- as expected
for an ILR.  It suggests the NIR bar is a second bar
inside the ILR, similar to nuclear bars in other galaxies (e.g. Friedli
et al. 1996).

\subsubsection{
\label{ecc20} 
Location of the Resonances}
\begin{figure}
\begin{center}
\psfig{file=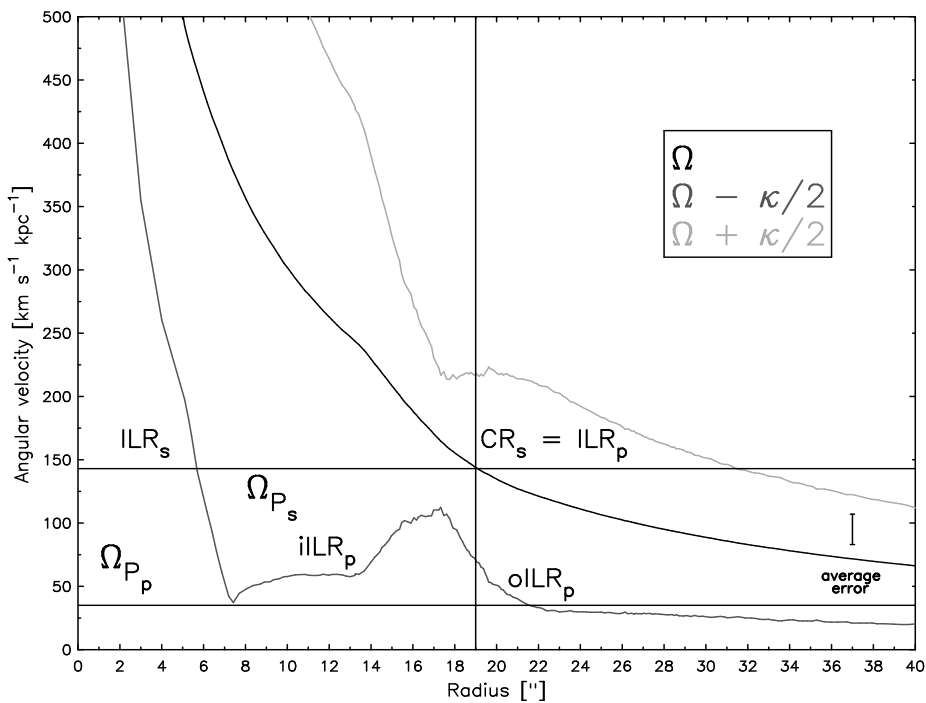,height=10.0cm,width=14.1cm,angle=-90.0}
\end{center}
\figcaption[]{
\label{vv8}
Sites of inner Lindblad resonances (ILRs) and corotation resonances
(CRs) in NGC~1068.
The ILR$_p$ of the outer, primary bar is at the molecular spiral arms
at $r \sim 18''$. If an inner, secondary bar exists, its CR$_s$
coincides with the ILR$_p$ of the outer bar. 
The outer, primary bar has a deprojected radius of about 120''
translating into a CR at about 140''. 
(See text.)
}
\end{figure}

We estimated the radii of the dynamical resonances of the inner and
outer bars from the rotation curve and bar lengths, using the curves
of $\Omega$ and $\Omega \pm \kappa/2$, to see if they matched the
sites of the gas ({\bf Fig.\,\ref{vv8}}).  (We follow the usual
nomenclature, with $\Omega =$ angular velocity, and $\kappa =$
epicyclic frequency, where
$\kappa^2 = r \times (d \Omega/d r)^2 + 4 \Omega^2 = 2 \Omega \times (\Omega 
+ (d v/d r)$ --- see, e.g. the review by Sellwood \& Wilkinson 1993). 

We approximated $d v/d r$ with the gradient $\Delta v/\Delta r$, which
means the derived curves of $\kappa$ and $(\Omega \pm \kappa/2)$ can
have errors due to the undersampling of the rotation curve.  The mean
error in the angular velocity at radii $>5''$ is $\pm$ 12\,\kms /kpc
and larger for radii $<5''$, where the velocity field is more
complicated. The local maximum in the $(\Omega - \kappa/2)$ curve at
$r \sim 17''$ corresponds to the maximum in the rotation curve.  The
change in the velocity gradient at this point results in a change in
$\kappa$, which yields the maximum in $(\Omega - \kappa/2)$.  The
change at this radius also occurs in the angular velocity curves of
Helfer \& Blitz (1995; their Fig. 18) and the stellar curves from
Telesco \& Decher (1988; their Fig. 6).

The outer or primary bar has a deprojected radius of 120''
(8.7\,kpc). At this radius, the HI rotation velocity is 165\,\kms\
(Brinks et al. 1997), so the pattern speed of the primary bar is
$\Omega_{P_p} \sim$ 20\,\kms /kpc with an uncertainty of $\pm
10$\,\kms /kpc because the end of the bar does not exactly coincide
with the corotation radius.  As shown in {\bf Fig.\,\ref{vv8}}, this pattern
velocity $\Omega_{P_p}$ yields an outer ILR (oILR) at a deprojected
radius of 1.6\.kpc ($17''$ on the sky) and an inner ILR (iILR) at
1.0\,kpc ($11''$ on the sky).  This is identical with the location of
the gas spiral arms at a deprojected distance of 1.3\,kpc ($\sim 14''$
on the sky).  It is thus likely that the gas spiral arms are caused by
the outer bar driving molecular gas into the resonance zone, where the gas 
becomes trapped.

If an inner bar exists, it should have its
corotation radius at the ILR of the outer bar.  This means the inner, 
NIR bar should have a deprojected radius of $\sim 1.4$\,kpc, as
observed. This radius yields an inner-bar pattern speed of 
$\Omega_{P_s} \sim $ 140\,\kms /kpc, which allows an ILR of
the inner bar at $r < 6''$ --- in the zone where we observe the
nuclear gas ring.  At these small radii, however, we know the
rotation curve and $(\Omega - \kappa/2)$ only poorly.  In
contrast to our ideas, Helfer \& Blitz (1995) estimated a deprojected
NIR bar length of only $15''$ and thus derived an pattern speed of
of $160$\,\kms . In their interpretation, the molecular spiral arms
lie between the outer Lindblad resonance and the corotation radius.

\section{MOLECULAR GAS MASS AND DYNAMICAL MASS}
We now discuss the molecular gas and dynamical masses for some
components of the \CO line emission. These values are used to estimate
the thickness of the gas disk (see section \ref{ecc7}) and the torques
acting on the disk, should it be warped (see section \ref{ecc3} and
Appendix C in SET99b).  Comparison of the molecular gas mass with the
dynamical mass of the nuclear region shows that the molecular gas is a
probe of the nuclear gravitational potential in NGC~1068, but not a
dominant component of that potential.

Since the interferometer maps may have missed up to one half the total line
emission, we derived masses only for the compact components (section
\ref{ecc1}).  The measured fluxes $S_{{\rm CO}}\Delta V$ [Jy\,\kms]
were converted into beam-diluted, integrated brightness temperatures,
in K\,\kms , and then multiplied by the Milky Way \nhico\ conversion
factor of $2 \times 10^{20}$ $\frac{{\rm cm}^{-2}}{{\rm K}\, {\rm
km}\,{\rm s}^{-1}}$ from Strong et al. (1989) to estimate H$_2$ column
densities.  We then multiplied by the beam-broadened source areas, in
cm$^2$, to obtain H$_2$ masses ({\bf Table \ref{vv42}}).  The
calibration errors are only 15\% (in the \COe line), so the main 
uncertainty in the masses comes from the \nhico-conversion factor,
which may be wrong by at least 50\% in the compact components, and
even more in the extended components. In these mass estimates, since
the integrated $\frac{{\rm CO}(2-1)}{{\rm CO}(1-0)}$ line ratio is
close to unity, we assumed the two CO lines were equivalent, provided
maps were convolved to the same resolution and intensities were
converted to K\,\kms .

We estimate the {\it dynamical mass}, in M$_\odot$, from $M_{\rm dyn}
= 232\ v^2(r)\ r$, where $v(r)$ is the inclination-corrected rotation
curve, in \kms , and $r$ is in pc.  The uncertainties in this mass
correspond to the uncertainties in the rotation curve (section
\ref{ecc4}).  Our value for the mass within $r = 2''$ (for circular
orbits) agrees with the upper limit from Thatte et al. (1997).  Most
of the gas in the inner $50''$ of NGC~1068 is in the molecular spiral
arms, and its mass is comparable with that in the molecular ring of
the Seyfert galaxy NGC~3227 (SET99b).  The molecular gas mass is only
5\% of the dynamical mass in the inner $50''$.  Even if one corrects
for CO emission resolved out by the interferometer, the gas
mass is less than 10\% of the dynamical mass.

\section{
\label{ecc7}
DISK HEIGHT FROM THE VELOCITY DISPERSION}
The velocity dispersion observed in the molecular line emission (see
{\bf Fig. \ref{vv3}{\it b} and {\it d}}) is due to (1) superposition of gas at
different locations on the line of sight with different velocities in
the same beam and (2) the intrinsic turbulent velocity dispersion of
single molecular clouds.  It is this turbulence that supports the disk and
determines its height.

\subsection{The velocity dispersion}
The velocity dispersion was measured in the second-order moment maps
of the \CO lines. In the regions of enhanced flux density on the
spiral arms the velocity dispersion rises to 20\,\kms\ whereas the
mean value in the spiral arms is $\sim$ 16\,\kms . This mean value is
similar to the HI velocity dispersion of 10\,\kms\ for radii $>$ 3~kpc
(Brinks et al. 1997). A higher HI dispersion of 30 to 50\,\kms\ is
observed at smaller distances from the nucleus (Brinks et
al. 1997). The discrepancy with the velocity dispersion measured in
\CO may be due to the larger HI beam ($8''$) at radii $< 20''$, or to
the HI disk being thicker than the molecular gas disk, as in the outer
regions of our Galaxy (Malhotra 1994, 1995).
For the bar region we also find a low dispersion of $\sim 16$\,\kms\
in the molecular gas. 

In the nucleus, the mean velocity dispersion measured in \COz is
40\,\kms , twice as high as in the spiral arms. This may be due to
inclination or beam smearing rather than intrinsic turbulence in the
gas.  A better estimate of the velocity dispersion of the nuclear gas
may be the 25\,\kms\ spread in the systemic velocity component in the
$pv$ diagram at p.a.\ 110$^\circ$.  This is probably the value that
sets the disk thickness.  The high apparent velocity dispersions in
the two nuclear emission knots are due to the superposition of at
least two distinct components (section \ref{ecc1}).

\subsection{The disk thickness}
To estimate the disk height from the velocity dispersion we must
correct for beam smearing. We assume the observed dispersion
$\sigma_{\rm obs}$ is the quadratic sum of the intrinsic dispersion,
$\sigma_{\rm real}$, and a contribution from the rotation curve
$\sigma_{\rm rot}$.
We adopt an apparent velocity dispersion of $\sigma_{\rm obs,sp} \sim
16$\,\kms\ for the spiral arms and $\sigma_{\rm obs,nuc} \sim
25$\,\kms\ for the nuclear region (see above). At the spiral arms, the
rotation curve is almost flat, so in this zone, the CO line widths are not
affected by beam smearing.
In the nuclear region, for $0.5'' < r < 1.0''$, 
the gradient in the rotation curve reaches 24\,\kms
/arcsec, which  in a $0.7''$ beam contributes $\sigma_{\rm
rot, nuc} \sim 10$\,\kms\ .  This only slightly increases the total
apparent velocity dispersion $\sigma_{\rm obs,nuc}$.  After deconvolution,
the intrinsic velocity dispersion in the nuclear region is
$\sigma_{\rm real,nuc} \sim 23$\,\kms .

There are several ways to relate the velocity dispersion $\sigma$ and 
disk height $h$. Assuming hydrostatic equilibrium they can be written
in their basic form as $h/r \sim \sigma/v_c$, with $v_c$ being the circular 
velocity at a radius $r$.
Quillen et al. (1992) and Combes \& Becquaert (1997) derived relations
for gas disks in the potentials of elliptical galaxies or galactic
bulges. Downes \& Solomon (1998) deduce a relation for disks with flat
rotation curves following the Mestel (1963) approximation. Both
approaches are relevant to the center of NGC~1068, where the potential
changes from disk-like to bulge-like.

For the {\it spiral arms}, with a rotation velocity $v(18'') \sim
200$\,\kms\ and a dispersion $\sigma_{\rm real,sp}\sim 16$\,\kms , the
equation 3.1 of Quillen et al. (1992) yields a disk height of 100\,pc.
For the {\it nuclear region}, with $v(1'') \sim 150$\,\kms\ and
$\sigma_{\rm real,nuc} \sim 23$\,\kms , we get the following
estimates. With eq.\ (3.1) of Quillen et al. (1992) we get a gas disk
thickness of 10\,pc.  With the (third) equation of Combes \& Becquaert (1997),
we find a disk height of 9\,pc.  The application of the Mestel
approximation as given by Downes \& Solomon (1998, their eq.\ 3) with
$\sigma_{\rm real,nuc} \sim 23$\,\kms , $v(3'') \sim 170$\,\kms\ and
$M({\rm H}_2) \sim 5.3 \times 10^7$\,\solm\ yields an estimate of
7\,pc. All three methods thus indicate the nuclear gas disk is thin
($d \sim 10$\,pc).  

\section{
\label{ecc3}
DYNAMICAL MODELING}
As noted in section \ref{ecc2}, the velocity field of the central
2.5\,kpc of NGC~1068 deviates from that of a simple inclined
rotating disk. An obvious deviation occurs at $r \sim 10''$, where the
isovelocity contour at the systemic velocity, which should be a
straight line along the kinematic minor axis, is in fact inclined by
45$^\circ$ to the kinematic minor axis of the outer disk.  Another
difference to a simple disk is the complex structure in the velocity
field and $pv$ diagrams of the inner few arcseconds.
Our kinematic models can fit these structures by either planar
elliptical orbits due to gas moving in bar potentials, or circular
orbits tilted out of the galactic plane in a warp.  We first describe
the best fits in pure bar models and pure warp models, and present a
combined solution as the most plausible explanation.  We then relate
this best-fit, mixed model to observations at other wavelengths.

\subsection{
\label{cc3}
Outline of our model program, 3DRings
}
Details of our 3DRings program are in SET99b (appendix B), so we give
here only a brief summary.  We assume that since gas is dissipative,
it persists only on orbits that are not self-intersecting, not
crossing, and not strongly cusped. Its orbits can thus be of only two basic
types: (1) planar elliptic orbits and (2) tilted circular orbits. The
first type are the $x_1$ and $x_2$ orbits along and perpendicular to
bars in the galactic plane. The second type are orbits that can leave
the galactic plane, thereby warping the gas disk.

The model divides the disk into many individual circular or elliptical
orbits of molecular gas.  For the modeling the inclination and
position angle of the disk, and the shape of the rotation curve were
held fixed.  Each fit was started at large radii and then extended to
the center. For each case we tried several setups that all converged
to similar best solutions with mean deviations from the data of $\leq
10$\,\kms\ and $0.1''$ for each velocity and radius in the
$pv$-diagrams and 10$^\circ$ in the position angle of the mapped
structures.

For the elliptical orbit models, we follow Telesco \& Decher (1988), 
who explained the gas motions between the two ILRs in NGC~1068 with a
smoothly varying position angle of the ellipses centered on the
nucleus.  In the bar models, we fitted the orbital eccentricity
$\epsilon (r) = b(r)/a(r)$ (where $a$ and $b$ are the major and minor
axes) and the position angle $PA_{\rm ellipse}(r)$ by curves varying
smoothly with radius, with the orbits constrained to be
non-overlapping.  The orbits lie in a plane and resemble the velocity
and density patterns in the bar models of Athanassoula (1992).

For the out-of-plane circular orbits, we follow the tilted ring model
outlined by Rogstad et al.  (1974) with a warp simulated by concentric
tilted rings.  The strength of the tilt and the precession of the
position angles are proportional to the acting torque. The torque can
result from gas moving in a non-axisymmetric potential, or from
interaction of the disk gas with the radiation pressure of a nuclear
source, with the gas pressure in the ionization cone of an AGN, or
with some GMCs out of the disk (see SET99b, Appendix C).

\begin{figure}
\begin{center}
\psfig{file=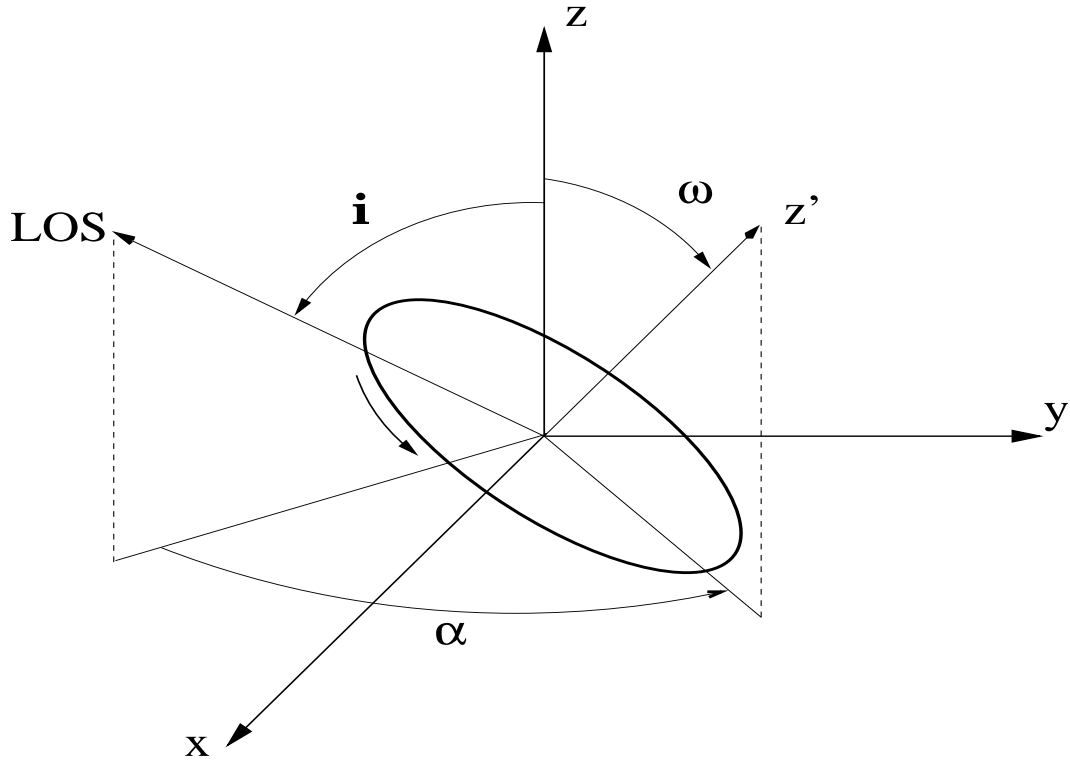,height=10.0cm,width=14.1cm,angle=-90.0}
\end{center}
\figcaption[]{
\label{vv20}
Parameters in the 3DRings warp model. 
The diagram shows only a single ring, tilted by an angle $\omega$
relative to the galaxy's rotation axis, which is inclined at angle $i$
to the line of sight.  Here $\alpha$ is the precession angle between
the projection of the line of sight on the plane
of the galaxy and the ring axis. 
}
\end{figure}

To model the warp, we followed the method of Tubbs (1980; see also
Quillen et al. 1992), in which the warp has a smoothly varying tilt
$\omega(r)$ of each orbit relative to the disk plane and a smoothly
varying precession angle $\alpha(r)$ ({\bf Fig. \ref{vv20}}). A torque
acting on an orbit with a circular velocity $v_c(r)$ induces a
precession rate $d \alpha/dt \sim \xi v_c/r$.  After a time
$\Delta$$t$ one obtains $\alpha(r) = \xi \Omega \Delta t + \alpha_0$,
where $\alpha_0$ is a constant, $\Omega$=$v_c$/$r$, and $\xi$ is given
by the acting torque.  Our models had constant $\xi \Delta t$ and
uniformly distributed gas.

%
The model fits were adjusted to match the observed $pv$ diagrams, the
velocity moment map, and the intensity map.  Comparison of the models
with the data was done in maps at a resolution of $0.4''$ which is
super-resolved relative to the synthesized beam .  This approach
allowed us --- especially in $pv$-diagrams --- to concentrate on the
brightest compact features in the data with a $0.7''$ beam.  For both
approaches we tried two rotation curves, one interpolated to 0\,\kms\
at $r=0''$ and one in which the observed rotation curve in the inner
few parsecs was replaced with a rotation curve of a fixed enclosed
mass. The best results were for a warp model with an enclosed mass of
$\sim 10^8$\,\solm .

\subsection{
\label{ecc6}
The pure bar approach
}
We showed above that two bars are clearly detected in NGC~1068 and
that the spiral arms are at the ILR of the outer bar (see section
\ref{ecc8}).  In the inner $4''$ the potential is dominated by the 
inner 1\,kpc NIR bar. 
To be stable, an inner bar must be weaker than an outer one
(e.g., Friedli \& Martinet 1993). The influence of the NIR-bar
on the gas motions is thus weaker than that of the outer
one (Macjiewski \& Sparke 1999). Furthermore, if the inner bar
controlled the gas motions, the large velocity dispersion observed at
a radius of $1''$ would be hard to explain, since strong streaming
motions are not expected for inner bars.  There is also no hint in the
NIR data (Thatte et al. 1997) for an additional, strong, {\it
third} bar that could influence the structure and kinematics of the
circumnuclear molecular gas.

\begin{figure}
\begin{center}
\psfig{file=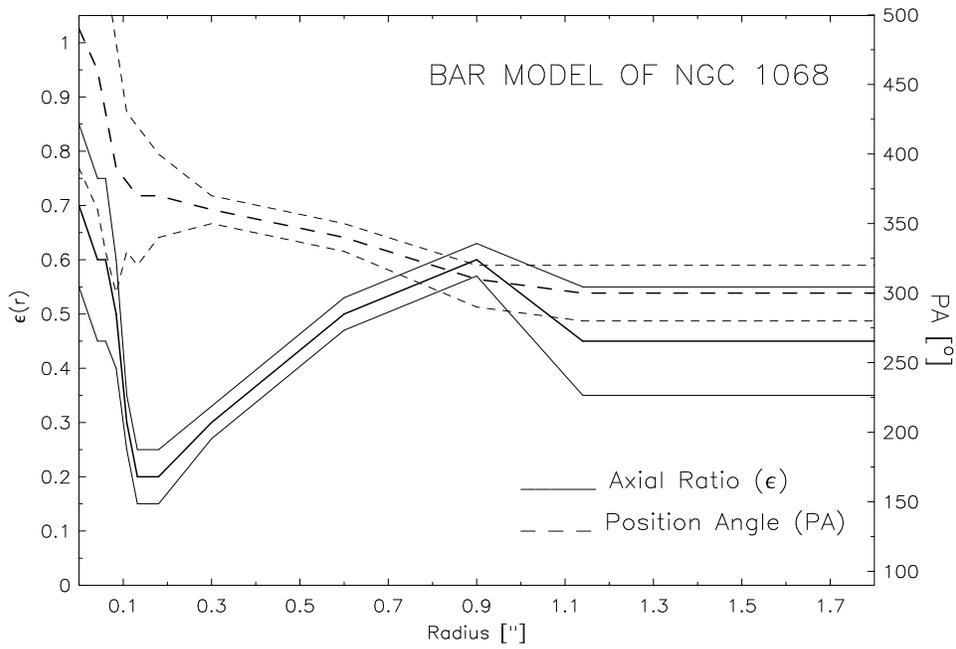,height=10.0cm,width=14.1cm,angle=-90.0}
\end{center}
\figcaption[]{
\label{vv9}
The curves of position angle $PA$ and ellipticity $\epsilon$ of
the elliptical orbits in the bar model that fit the data.
The curves of $\epsilon$ (solid line) and $PA$ (broken
line) are shown as thick lines for the best fit, 
and as thin lines over the range of satisfactory fits (see
text).  
}
\end{figure}

\begin{figure}
\begin{center}
\psfig{file=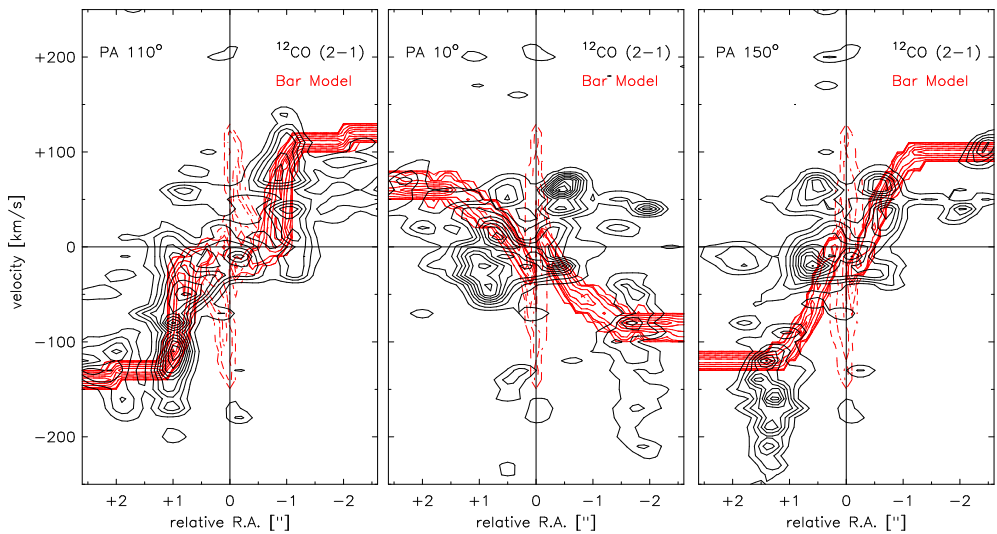,height=10.0cm,width=14.1cm,angle=-90.0}
\end{center}
\figcaption[]{
\label{vv10}
Position-velocity diagrams for the bar model in NGC~1068. 
The regions of high velocity dispersion that appear at low intensity 
(p.a.\ 10$^\circ$, radius $-1.8''$ and p.a.\ 150$^\circ$, $r=1.3''$) are 
molecular cloud complexes that are  not part of a symmetric
velocity field (see section \ref{ecc1}). The bar model explains  
all symmetric structures. Although
the bar model can fit the component at lowest velocities, 
it cannot explain the high velocity dispersion of
these components.  The data at $0.4''$ resolution are shown in
black contours of 10 to 100\% in steps of 10\%. The model,
calculated for a starting radius of 27\,pc, is shown in red
contours at 0.2, 2, 5, 20, 40, 60 and 80 \%.
}
\end{figure}

\begin{figure}
\begin{center}
\psfig{file=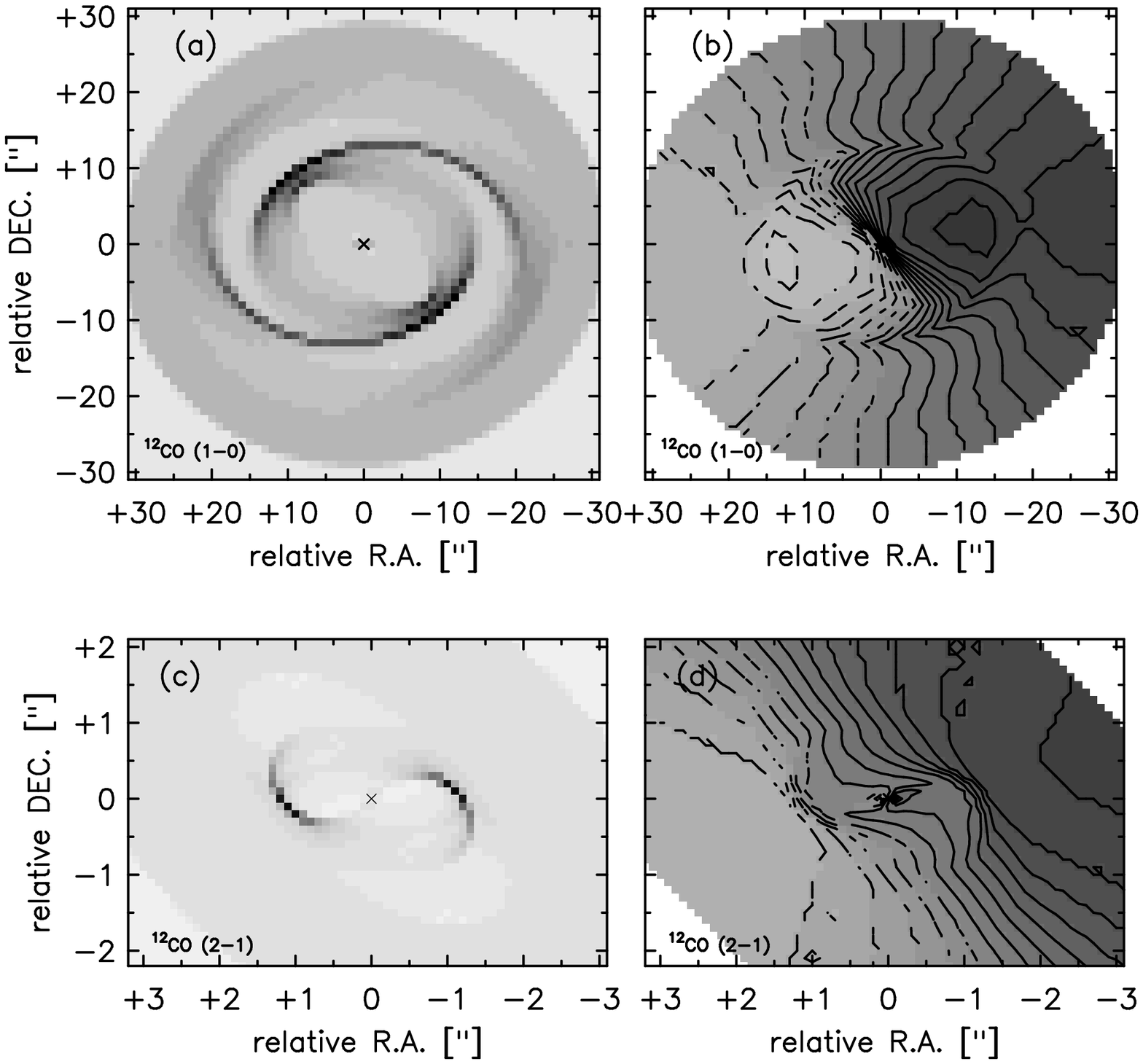,height=14.1cm,width=10.0cm,angle=0.0}
\end{center}
\figcaption[]{
\label{vv11}
Intensity maps ({\it left}) and velocity fields ({\it right}) of the bar model
for the \CO emission ({\it a and b}: \COe, {\it c and  d}: \COz). 
Velocity contours are from $-$150 to 150\,\kms\ in steps of
20\,\kms\ for \COe and from $-$130 to 130\,\kms\ in steps of 20\,\kms\
for \COz. Broken lines indicate negative velocities.
}
\end{figure}

For our pure bar model, 
{\bf Fig.\,\ref{vv9}} shows the model curves for the position angle and
ellipticities of the fitted ellipses, with their errors (see section
\ref{cc3}).  {\bf Figure\,\ref{vv10}} compares the model $pv$ diagrams with
the data, and {\bf Fig.\,\ref{vv11}} shows the model intensity distribution
and velocity field.  We now discuss the different regions of the model.

{\it Spiral arms and bar region:}
Our best bar model gives a good fit to the data cube in the spiral
arms and in the transition to the NIR bar ({\bf Fig. \ref{vv11}}).
In particular, the model reproduces the shifts in the alignment of the
kinematic minor axis.  Inspection of the velocity field on the
northern part of this axis reveals the behavior of the molecular gas
at an ILR and near the inner bar. At an ILR (the spiral arm region in
NGC~1068), the velocity is blueshifted relative to systemic, and
becomes redshifted near the NIR bar.  That is, at the $30''$ diameter
spiral arm region, the velocity on the minor axes in both \COe and
\COz is blueshifted relative to systemic because of the ILR of the outer
large-scale bar.  This
velocity change along the minor kinematic axis occurs close to the 
spiral arms, at the transition
from the $x_1$ orbits (those running along the outer bar, outside the
oILR) to the $x_2$ orbits (those normal to the outer bar, between the
oILR and the iILR). This outer velocity change is due to the $x_2$
orbit velocities being high where the $x_1$ orbit velocities are low.
\\
At radii $\leq 7''$, the velocity is redshifted because of the NIR
bar.  At the iILR, there is a transition from $x_2$ orbits of the
outer bar to $x_1$ orbits of the inner, NIR bar at p.a.\
48$^\circ$ (Thronson et al. 1989).  This is the second, inner,
velocity change along the kinematic minor axis.  In NGC~1068 this
second shift is seen in the \COe and $^{13}$CO(1--0) lines (see
section \ref{ecc10} and Tacconi et al. 1997; Helfer \& Blitz 1995).
This scenario is well matched by the orbit crowding in our model.

{\it The nuclear gas ring:}
Near the nuclear ring, the bar model fails to reproduce the observed
velocity increase moving in to $1''$, the subsequent abrupt drop to
the systemic velocity, and the observed ridge between the two CO knots
(section \ref{ecc1}; {\bf Fig.\,\ref{vv4}}).  This is a problem for
the pure bar model, and one may ask if it could be rescued with a
third bar.  
Fitting of the gas motions with the bar approach suggests the presence
of highly elliptical east-west orbits and therefore provides evidence for the
the presence of a third bar with a major axis of $1''$ length.
High-resolution NIR
observations, however, show no such nuclear bar in NGC~1068
(Wittkowski et al. 1998; Weinberger et al. 1999; Thatte et
al. 1997). The stellar continuum in the nucleus has neither the $1''$
size nor the shape expected for a third bar (see Fig.2 in Thatte et
al. 1997).

\subsection{The pure warp approach}
Another model for non-circular motions is a warp in the gas disk.  The
warp model yields a solution that is fully consistent with the derived
gas {\it rotation curve}.  In particular, it can explain the high
velocities between $1''$ and $2''$ in the nuclear ring, seen in the
$pv$ diagram at p.a.\ 110$^\circ$.  (As in the bar model, for the
inner few parsecs we adopted a Keplerian rotation curve for an
unresolved point mass --- cf.\ section \ref{cc3}.)

The {\it precession} of the warp is given by the torque $\xi$, the
time interval $\Delta t$, the initial precession angle $\alpha_o$, and
the angular velocity $\Omega(r)$ of the rotation curve.  For the
best model, we get $\alpha_o = 10^\circ \pm 15^\circ$ and $\xi \Delta
t = (7.0\pm0.5) \times 10^5$ yr. In this model, it is the fit to the
intensity distribution that is more sensitive to changes of $\xi
\Delta t$, not the fits to the $pv$ diagrams. The precession time is
much longer than the rotation time at $r = 1''$. At this
radius, with our preferred value of $\xi \Delta t$, the disk must make 
25 rotations in order to precess by 360$^\circ$.

\begin{figure}
\begin{center}
\psfig{file=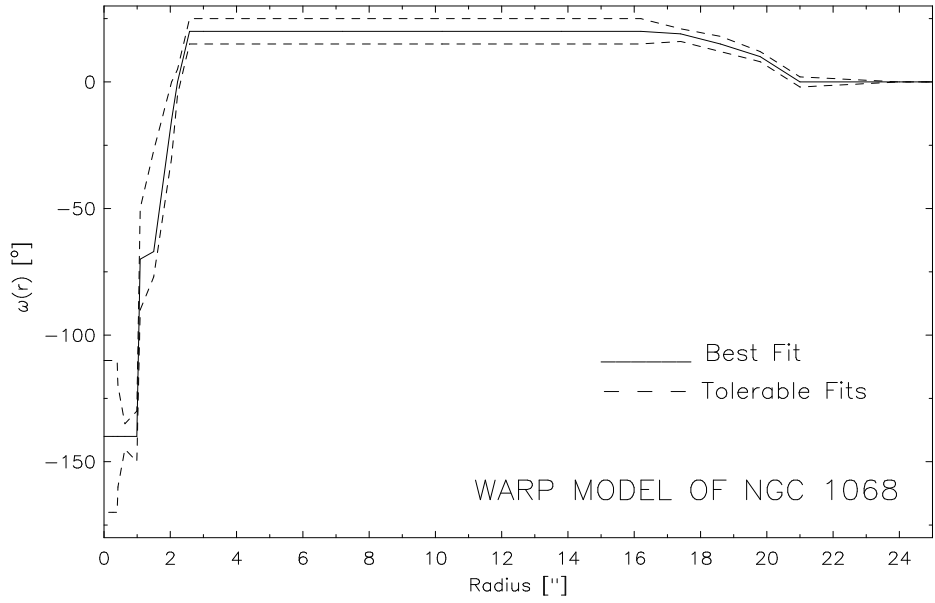,height=10.0cm,width=14.1cm,angle=-90.0}
\end{center}
\figcaption[]{
\label{vv12}
The $\omega$ (angular velocity) curve of the warp model. 
The solid line shows the best fit to the data.  Broken lines indicate
the range over which the fits are satisfactory.
}
\end{figure}

The inclination curve $\omega(r)$, which is the {\it tilt of the
warp}, is shown in {\bf Fig.\,\ref{vv12}} for the best fit.  The errors range
from $\pm$ 5$^\circ$ to $\pm$ 30$^\circ$. The small errors at $r =
1''$ are due to the abrupt drop in the systemic velocity at this
radius, which in turn depends strongly on the inclination of
individual rings.

{\it The spiral arms and bar region:} 
A warp can mimic spiral patterns in the plane of the sky
(Steinman-Cameron et al. 1992).  What about the molecular spiral arms
in the center of NGC~1068?  Their kinematic axis changes direction,
suggesting these arms depart from circular rotation and may also be a
warp effect.  We found that to fit the spiral arms, the model disk
must change its inclination by 20$^\circ$ between $r= 20''$ to $17''$,
then stay at this new inclination down to $r \sim 2.6''$, and then
flip back to the old inclination at $r = 2.2''$.  One anomaly is the
kinematic minor axis, which in the observed data, crosses the
zero-velocity contour twice, at the spiral arms.  The warp model
cannot reproduce this effect.  Otherwise, the warp model fits the
velocity field inside $r=17''$ just as well as the bar model does.

\begin{figure}
\begin{center}
\psfig{file=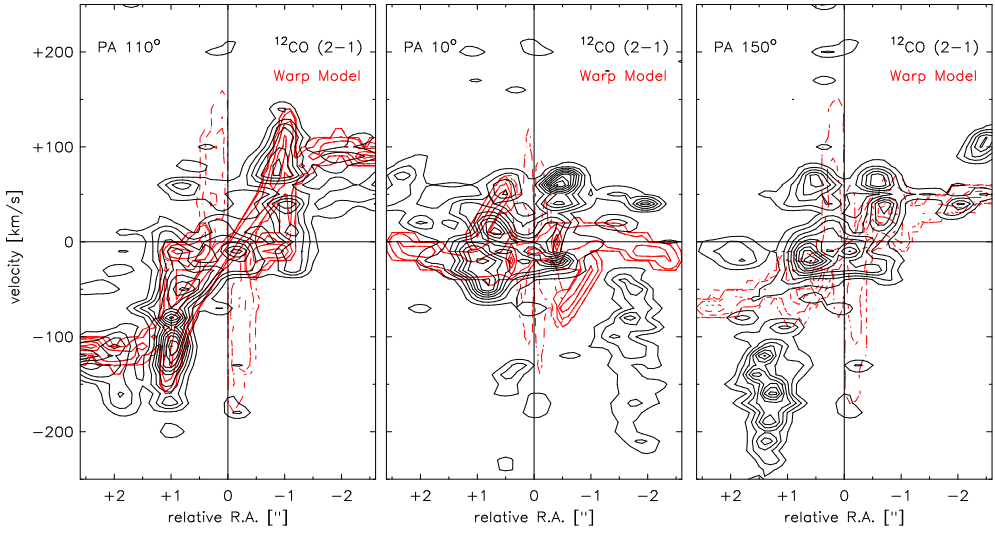,height=10.0cm,width=14.1cm,angle=-90.0}
\end{center}
\figcaption[]{
\label{vv13}
Position-velocity diagrams for the warp model in NGC~1068. 
The weak emission regions of high velocity dispersion 
(at p.a.\ 10$^\circ$, $r = -1.8''$ and p.a.\ 150$^\circ$,  $r=1.3''$)
belong to molecular cloud complexes which are not part of a symmetric
velocity field (see section \ref{ecc1}). Our model only explains
symmetric structures. The contrast-enhanced data are in black contours
of 10 to 100\% in steps of 10\%. The model, calculated
starting from $r = 27$\,pc, is shown in red contours at 0.2, 3, 20, 50 
and 80 \%.
}
\end{figure}

\begin{figure}
\begin{center}
\psfig{file=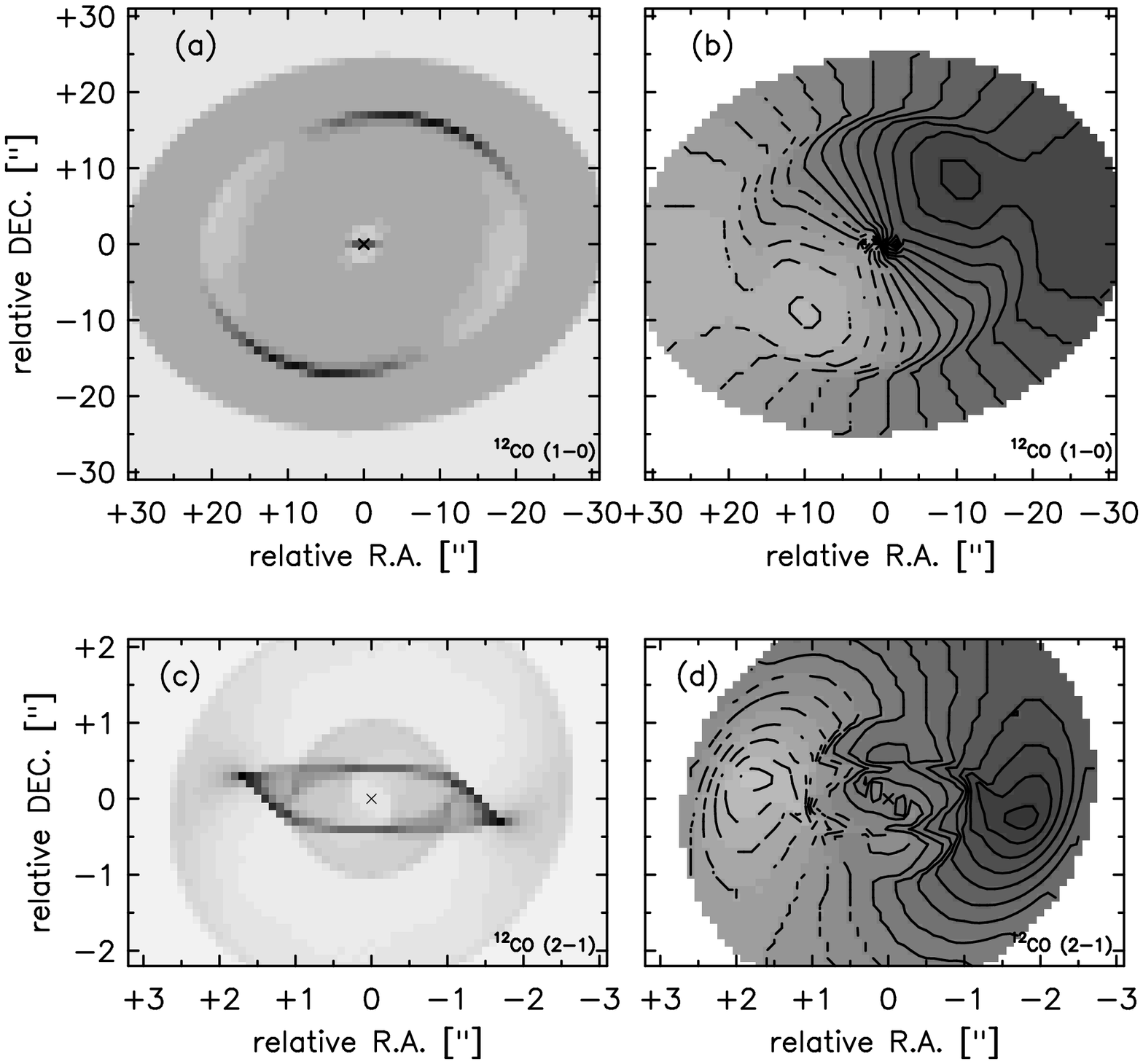,height=14.1cm,width=10.0cm,angle=0.0}
\end{center}
\figcaption[]{
\label{vv14}
Intensity maps ({\it left}) and velocity fields ({\it right}) of the warp model 
for the \CO emission ({\it top:} \COe, {\it bottom:} \COz). 
Velocity contours are from $-$150 to 150\,\kms\ in steps of 20\,\kms\
for \COe and from $-$130 to 130\,\kms\ in steps of 20\,\kms\ for
\COz. Broken lines indicate negative velocities.
}
\end{figure}

\begin{figure}
\begin{center}
\psfig{file=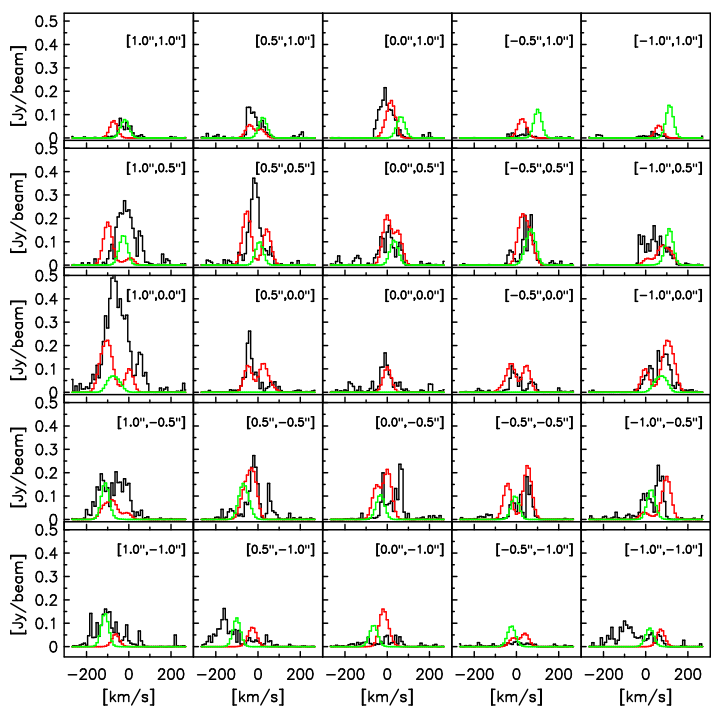,height=14.1cm,width=10.0cm,angle=0.0}
\end{center}
\figcaption[]{
\label{vv14a}
Comparison of measured spectra ({\it black}) with those given by the warp
model ({\it red}) and the bar model ({\it green}), in the central $2'' \times 2''$
in steps of $0.5''$.
Intensity scales are arbitrary. With the exception of the
eastern knot (knot E) the warp model represents an acceptable fit to
the observed data that reflect the complex velocity field. The agreement
of the bar model with the data is not
as good as it cannot reproduce the multiplicity (two features) of the
line, especially at positions west and south-west of the nucleus.
}
\end{figure}

{\it For the nuclear ring,}
{\bf Fig.\,\ref{vv13}} compares the observed and
model $pv$ diagrams, and {\bf Fig.\,\ref{vv14}} shows the intensity
map and the velocity field of the warp model.  For the symmetric $pv$
diagram at p.a.\ 110$^\circ$ with most of the line flux, the model
deviates from the data by only $\pm 5$\,\kms . In the nuclear ring,
the model can explain everything in the $pv$ diagram and the spectra
({\bf Fig.\,\ref{vv14a}}): the large
velocity dispersion, the associated increase in velocity at $1''$, the
emission at the systemic velocity at $r < 1''$, the decrease of
velocity toward the center, and the ridge between the two high
dispersion points. On the minor kinematic axis, the fit is not as
good, with deviations of $\pm 20$\,\kms . This may be due to the low
line flux and its contamination by sidelobes that are not completely
removed by the CLEAN algorithm.

For radii $< 2.2''$ the disk starts warping, down to $r \sim 1.5''$
(110\,pc) where it becomes almost edge-on ($i'=i+\omega \sim
70^\circ$). It stays at this inclination for about 35\,pc ($0.5''$)
toward smaller radii, and then warps within $r<10$\,pc from edge-on to
almost face-on.  Due to resolution limits, the orbit orientations at
$r < 30$\,pc are not well defined. The outer warped molecular gas disk
(10\,pc to 100\,pc) probably joins continuously the inner accretion
disk ($\sim 1$\,pc), which may also be warped (Gallimore et al. 1997;
Greenhill \& Gwinn 1997). The warped molecular gas disk may thus be
not only the link between the host galaxy and the accretion disk, but
also the fuel reservoir for the AGN (see section 8).

\subsection{\label{ecc11}The mixed model}
The bar and warp approaches both provide good fits to the data for
specific regions of NGC~1068. From the observations of the two bars
and the analysis of the dynamical resonances, it is obvious that the
modeling of the spiral arms with elliptic orbits representing a bar
potential is plausible and physically realistic.  This is not
necessarily true for the nuclear region, since the potential of the
inner stellar (NIR) bar is weaker relative to the outer bar, and
the warp approach provides a better fit to the individual components.

\begin{figure}
\begin{center}
\psfig{file=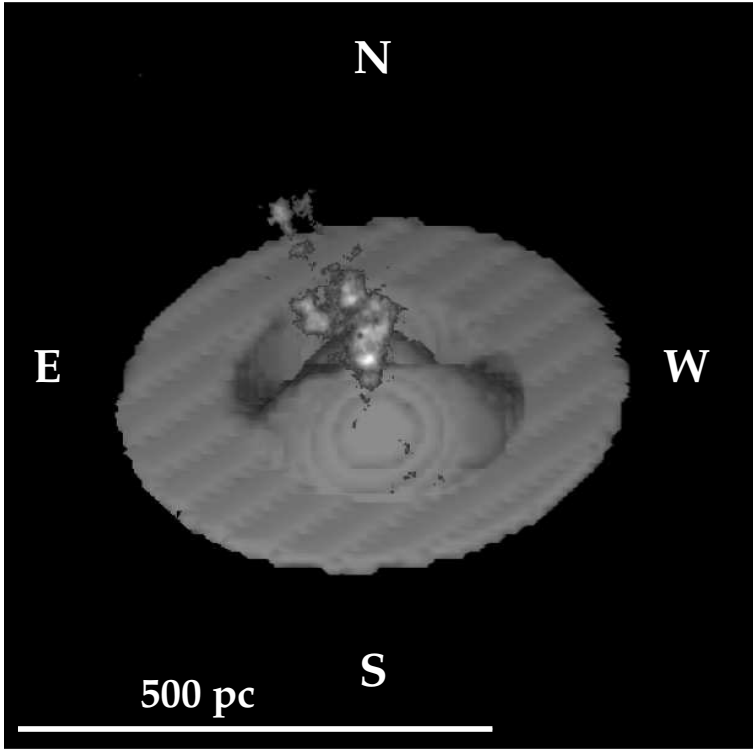,height=10.0cm,width=10.0cm,angle=0.0}
\end{center}
\figcaption[]{
\label{vv15}
Geometry of the NGC~1068 warp model. 
Brighter sections are closer to the observer, darker ones are farther
away.  The brightest structure is the [O~III] ionization cone
(Macchetto et al. 1994).  The 3-dimensional geometry of the warp
creates a natural cavity for the ionization cone that is consistent
with its observed orientation.
}
\end{figure}

We thus use a hybrid model to describe the kinematics of NGC~1068,
combining the best fits from both approaches.  In the warp model, the
tilt of the nuclear disk goes in the opposite direction of that needed
to model the spiral arms. This allows us to join, at the outer edge of
the nuclear ring, the warp model for the inner zone with the bar model
for the outer zone.  This hybrid model fits very well the intensity
and velocity distributions of all the symmetric features.
{\bf Figure\,\ref{vv15}} is a perspective view of the 3-dimensional model
in the plane of the sky.  In this mixed model, the spiral arms are
at the ILR of the outer bar. The NIR bar is an inner bar
with its $x_1$ orbits causing the change in the position angle
inside the spiral arms.  The bar model of both the ILR of the outer
bar (at the spiral arms) and the transition to the inner bar fits well
the observed velocity field.  The nuclear ring, however, cannot be
reproduced very well as an ILR of the inner bar, in the same plane.
The warp approach gives a better fit to the nuclear gas ring.

\section{
\label{ddd}
DISCUSSION
}
In section \ref{ecc8} we described the 17\,kpc-long outer bar and
the 2.7\,kpc-long, NIR inner bar.  Gallimore et al. (1999)
recently resolved the HI absorption in the nuclear region of NGC~1068
in both space and velocity.  The strongest HI is $4''$ southwest of
the nucleus and absorbs continuum emission of the southern radio lobe 
(Fig.\,4 of Gallimore et al. 1999). The HI absorption velocity varies from
southeast to northwest.  Gallimore et al. (1999) modeled this motion
with a disk inclined at the position angle of the stellar bar, and
suggested the motion may be due to gas in the potential of that bar.
The velocity fields of our \COe data and our bar model confirm this conclusion.
Comparison of systemic velocity measurements (Helfer \& Blitz 1995;
Brinks et al. 1997; Greenhill \& Gwinn 1997; this paper) suggests the
value for NGC~1068 is $\sim 1120$\,\kms\ rather than the 1150\,\kms\
derived by Brinks et al. (1997) from their HI data with a $8''$
beam. This implies that the HI absorption at 1110 and 1131\,\kms , seen in
projection near the nucleus in the maps of Gallimore et al. (1999), is
actually physically associated with the nucleus.

Friedli \& Martinet (1993) made N-body simulations of double-barred
galaxies and showed the dynamics inside the outer bar's ILR can be
decoupled, so that an inner bar can form. This is consistent with
our bar model for the \CO data.
The inner and outer bars are usually assumed to
have similar ILRs (Friedli \& Martinet 1993; Maciejewski \& Sparke
1999).  So far, however, simulations of inner-bar ILRs have not
been done with high enough numerical resolution to yield the
details of such a region.

For the outer bar to survive, the inner bar's potential must be much
weaker, implying the inner bar has a less pronounced ILR than the
outer bar (Friedli \& Martinet 1993).  The molecular gas ring at a
radius of 75\,pc from the nucleus of NGC 1068 is close to the central
stellar cluster (Thatte et al. 1997). The gravity of the central mass
concentration competes with that of the inner, NIR bar, further
weakening the inner bar's influence on the gas in the ILR region.
Near the inner bar's ILR, a vertical resonance may evolve (e.g. Combes
et al. 1990).  Since only non-crossing orbits can retain the gas, such
a vertical resonance would warp the inner molecular disk. Once out of
the disk, the gas would be further supported by the potential of the
central stellar cluster, competing with that of the bar.

{\it Advantages of the warp model for the nuclear region:}
In summary, the reasons favoring a warp model are: 
(1) The best warp model reproduces all the observed symmetric
structures in the CO $pv$ diagram along p.a. 110$^\circ$.
(2) The warp forms an opening for the ionization cone 
({\bf Fig.\,\ref{vv15}}).
(3) The warp provides enough dust between the observer and the AGN to
explain why the NIR peak is offset from the optical one (see
below).
(4) In the inner $3''$, the polarization vectors are aligned as expected
for the edge-on geometry of the molecular gas in the warp model (see
below; an edge-on geometry for the central few arcsec was also
proposed by Baker \& Scoville (1998) and Baker (1999) to explain their
CO data).
  
{\it Cause of the warp:}
To move gas out of the plane, a torque is needed (see appendix C in SET99b). 
Torques can be
induced by a non-spherical galactic potential (just as the halo acts
on the HI disk), by radiation pressure from a radio jet (similar to
central radiation sources causing warps in accretion disks; Pringle
1996; 1997), by gas pressure in an ionization cone (Quillen \& Bower
1999; SET99a,b) or by the transient gravitational force of a giant
molecular cloud temporarily dislocated above the plane of the disk.
Arnaboldi \& Sparke (1994) calculated the torque of a slightly oblate
galactic potential outside the core.  We assumed a similar potential,
and used their eq.\,(B7) to estimate the torque.  Our results ({\bf
Table \ref{vv44}}) show the most likely causes of a warp are torques
induced by gas pressure, or by dislocated GMCs.  The torque induced by
the gas pressure of the ionization cone, for example, could persist
and maintain the warp for $\sim 10^8$ yr.  A warp caused by a GMC
(e.g., the western knot), deviating from the general rotation, would
only perturb the galactic potential for as long as the GMC is above or
below the molecular disk.

\begin{figure}
\begin{center}
\psfig{file=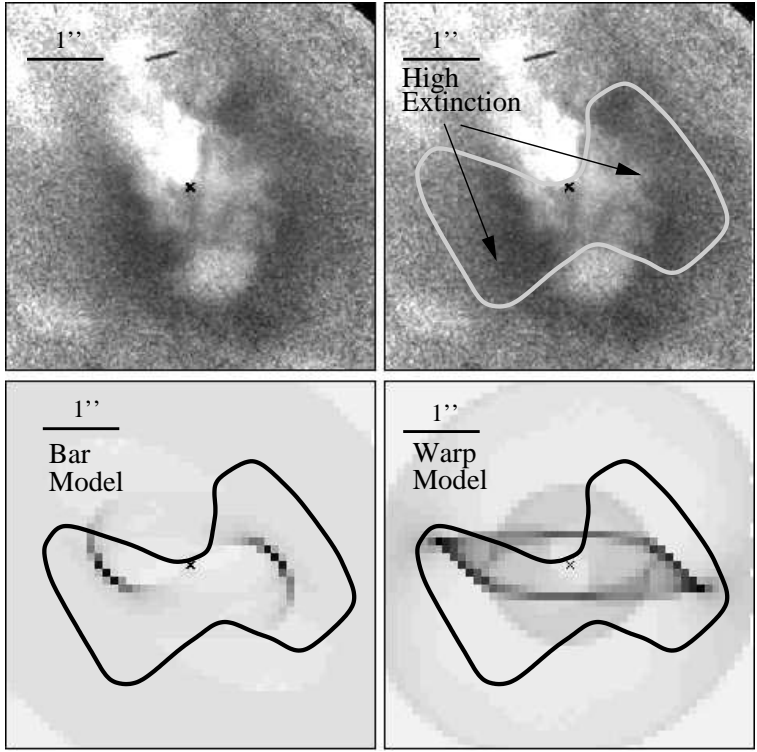,height=10.0cm,width=10.0cm,angle=0.0}
\end{center}
\figcaption[]{
\label{vv16}
The F550M filter HST image ({\it top left}) of the central $4'' \times 4''$
in NGC~1068 divided by a model of the stellar light 
(see text and Catchpole \& Boksenberg 1997). 
The dust lane is encircled by the solid gray line in
({\it top right}). The {\it bottom right} image shows that the CO 
intensity predicted by the warp model agrees with the dust
distribution. The agreement is not as good for the bar model ({\it bottom
left}), which does not predict the gas and dust
concentration above and below the nucleus.
}
\end{figure}

{\it Optical dust lanes:}
The dust lanes in the center of NGC~1068 are obvious in the HST image
with the F550M filter (Fig.\ 3 of Catchpole \& Boksenberg 1997). To
enhance the contrast, the authors subtracted a model of the nuclear
starlight from the image. Their model was azimuthally symmetric,
following two power laws. The model-corrected HST image nicely shows
enhanced extinction running east-west.  {\bf Figure \ref{vv16}} compares this
corrected image with the intensity distributions from the bar and warp
models. This comparison favors the warp model, which predicts
absorbing material at about the right locations. In a planar bar
model, the extinction in front of the nuclear region can be explained
only by postulating a very thick nuclear disk, which is inconsistent
with disk thicknesses estimated from the velocity dispersion of the
molecular gas.

{\it The near- and mid-IR nuclear polarization:}
The observed near- and mid-IR polarization vectors in the inner $3''$
are aligned east-west (Packham et al. 1997; Young et al. 1996).  For
Mie scattering in an edge-on disk, light should be polarized parallel
to the disk.  The reason is as follows: Bands of parallel polarization
vectors are often observed in bipolar stellar outflows and can be
reproduced by multiple scattering models (Bastien \& Menard 1988;
Fischer, Henning, \& Yorke 1996). These bands are often normal to the
flow and occur especially in edge-on disks with high densities and
particle sizes that are large relative to the wavelength of scattered
light (Fischer et al. 1996). Besides strong forward and backward
scattering, the Mie process also yields highly polarized emission,
with the $E$ vectors and the maximum polarized intensity both
perpendicular to the direction of forward scattering. The light is
strongly concentrated in two outflow cones, with the intense,
back-scattered light illuminating the disk. This scattered light is
then polarized parallel to the disk, over the total illuminated area
(Fischer et al. 1996; Bastien \& Menard 1988).  Bands with parallel
polarization thus occur most clearly in edge-on disks.

From the near- and mid-IR polarization maps of NGC~1068, Young et
al. (1996) and Packham et al. (1997) derive scattering cones at p.a.\
30$^\circ$ and a ``torus'' size of 200\,pc ($3''$), consistent with
the orbit diameters in our warp model, for the orbits with large
inclinations to the line of sight.  Alternatively, the alignment of
the NIR polarization vectors could be due to absorption of the
central stellar continuum by elongated dust grains aligned in a
magnetic field by the Davis-Greenstein effect (Greenberg 1978). If so,
then the magnetic field lines should lie in the plane of the absorbing
gas disk, parallel to the polarization vectors.  In this scenario,
however, the mid-IR polarization vectors should be orthogonal to the
NIR vectors, since the scattering efficiency is low at long
wavelengths (unless the grains are very large), and the mid-IR flux is
dust emission rather than scattered light.

\section{CONCLUSIONS}
\label{eee}
1. {\it Molecular gas close to the nucleus.---} Our new
interferometric data allow a high-sensitivity study of the molecular
gas in the inner 1~arcmin of NGC~1068 at high angular ($0.7'' \times
0.7''$) and spectral resolution (10\,\kms ).  We made detailed maps of
the 200\,pc-diameter nuclear ring indicated in previous molecular-line
maps (Tacconi et al. 1997).  We now find indications of molecular gas
emission as close as 13\,pc from the nucleus with a large offset from
the systemic velocity.  This emission suggests a mass within the inner
25\,pc of 10$^8$\,\solm . This is consistent with the black hole mass
of $1.7\times 10^7$\,\solm\ indicated by the H$_2$O masers (Greenhill
\& Gwinn 1997), plus a significant contribution from the mass of the
nuclear stellar cluster (Thatte et al. 1997).

2. {\it Bar and warp model.---}
Modeling of the molecular gas kinematics in the inner 50'' 
suggests a hybrid model combining bar-induced motion in the spiral arms and 
NIR bar with a warp of the thin gas disk in the 
nuclear ring (inner 300~pc). Due to the warping, the molecular gas  
leaves the plane of the disk at a radius of about
150~pc and has an inclination of $\sim 90^\circ$ to the line of
sight at about 75~pc.

3. {\it Extinction of the nucleus.---}
The obscuration of the AGN by the warped molecular gas disk is
consistent with the observations of the near- and mid-IR polarization
and the observed extinction of the nucleus at optical wavelengths. It
explains its identification as a Seyfert~2 galaxy with a hidden
Seyfert~1 nucleus. 

4. {\it Cause of the warp.---}
The probable cause for warping of the gas disk is the gas
pressure in the ionization cone. This scenario is also proposed for
the geometry in M~84 (Quillen \& Bower 1999). Further possibilities
are the influence of the transition between the disk/bar and bulge
potential, or dislocated GMCs transiently disturbing the gas disk.

5. {\it Small tori are not needed.---} 
As the results by Gallimore et al. (1997) show it is very likely that
NGC~1068 has a $<$2~pc molecular torus.
However, our findings imply that the nuclear molecular gas disks may be warped
providing plenty of material to occult the nuclear regions.
Therefore we do speculate that long-postulated
(large-scale) molecular tori with radii of 
a few 10 pc to a few 100 pc are not needed in every case.
The more likely explanation of the Seyfert
1/Seyfert 2 differences is that the molecular gas in the center of the
host galaxy contributes significantly to the obscuration of the active
nucleus and thereby influences the classification (Malkan et al.
1998). In our model, the absorbing molecular clouds need no longer be
confined to the galaxy's plane, in agreement with the geometry already
proposed by Cameron et al. (1993) for NGC~1068. This implies that the
apparent nuclear properties are strongly influenced by the distribution
of gas and dust in the center of the host galaxy.

6. {\it The gas spiral arms are at the ILR of a large-scale outer bar.---}
Analysis of the structure of NGC~1068 shows that the 3\,kpc
long, NIR stellar bar is an inner bar, since the galaxy also has
an outer bar-like structure about 17\,kpc long. This means that the
spiral arms in the molecular gas (at a radius of 1.4~kpc) are at the
ILR of the outer bar, in agreement with the theory of gas dynamics in
barred potentials. Comparison of the velocity fields of the molecular
gas and the stars with our model velocity field of a bar shows they
are remarkably similar.

\acknowledgements
IRAM is financed by INSU/CNRS (France), MPG (Germany) and IGN (Spain).
 We thank the staff on Plateau de Bure for doing the observing, and the
 staff at IRAM Grenoble for help with the data reduction,
 especially R.\ Neri, S.\  Guilloteau and J.\ Wink. For fruitful 
 discussions we thank R.\ Antonucci, A.\ Baker,
 P.\ Englmaier, J.\ Gallimore, O.\ Gerhard, R.\ Maiolino, A.\ Quillen,
 N.\ Scoville, L.\  Sparke and N.\ Thatte. We used 
 the NASA/IPAC Extragalactic Database (NED) maintained by the
 Jet Propulsion Laboratory, California Institute of Technology, under
 contract with the National Aeronautics and Space Administration.

\newpage


\clearpage

\clearpage

\clearpage

\clearpage

\clearpage

\clearpage

\clearpage

\clearpage

\clearpage

\clearpage

\clearpage

\clearpage

\clearpage

\clearpage

\clearpage

\clearpage

\clearpage

\clearpage

\begin{table}[htb]
\caption{
\label{vv40}
Properties of NGC~1068
}
\begin{center}
\begin{tabular}{lc}\hline \hline
 &  NGC~1068 \\ \hline 
Right ascension (J2000) &  02$^{\rm h}$ 42$^{\rm m}$ 40.798$^{\rm s}$\\
Declination (J2000) &  $-00^\circ$ 00$'$ 47.938$''$\\
Classification & (R)SA(rs)b \\
Inclination &  40$^\circ$\\
Position angle & 278$^\circ$\\
AGN Type &  Sey~2\\
Systemic velocity & 1150\,\kms \\
Distance &  14.4\,Mpc\\
\hline \hline
\end{tabular}
\end{center}
Nuclear coordinates are from Muxlow et al. (1996). Systemic 
velocity and Seyfert type are from NED (NASA/IPAC Extragalactic Database). 
The host galaxy classification is from de Vaucouleurs et al. (1991). 
Inclination and position angle are 
from the ``Ringberg standards'' (Bland-Hawthorn et al. 1997).
\end{table}

\begin{table}[htb]
\caption{
\label{vv42} 
Molecular gas masses of various compact components in NGC~1068
}
\begin{center}
\begin{tabular}{lrrr}\hline \hline
component 	& $S_{\rm CO}\,\Delta V$ 
		& $M$(H$_2$) 
		& M$_{\rm dyn}$ 
\\
 		& [Jy km/s] 
		& [10$^{7}$ M$_{\odot}$] 
		& [10$^8$ M$_{\odot}$] 
\\ 
\hline
total & 670 &  68 & 120 \\
spiral arms & 560 &  57 & \\
Northern bar & 2140 &  5 & \\
ring in \COz & 20 &  5 & 9\\
knot $E$ & 12 &  0.3 & \\
knot $W$ & 8 &  0.2& \\
\hline \hline
\end{tabular}
\end{center}
Uncertainties in gas masses are 40\% for the compact
components and 50 \% for the extended components. 
Uncertainties of dynamical masses are dominated by rotational velocity errors. 
The fluxes were derived for the total line
width with exception of the two knots in the ring for which we used
only velocities from $-230$ to $-30$\,\kms . The flux of knot $E$ was
measured in a rectangle with corners at $-1.3''/-2.0''$ and
$0.6''/-0.9''$ and knot $W$ in a rectangle with corners at
$0.8''/-1.3''$ and $1.9''/-0.4''$.  The dynamical mass was derived
with the equation in the text, with our molecular gas rotation curve
at radii of $18''$ and $2''$.
\end{table}

\clearpage
\notetoeditor{The following two tables should be one;
     table caption is incomplete in processed file (ps-file)}
\begin{table}[htb]
\caption{
\label{vv44}
Causes for the warp in NGC~1068
}
\begin{center}
\begin{tabular}{lllr}\hline \hline
Cause & Property & Value \\ \hline
Estimated via & & \\
{\bf 3DRings}$^a$   & {\bf torque}$^1$ & $M$  &{\bf  1.9$\times$10$^{47}$ Nm}  \\
                & molecular gas mass   & $m$  & 2.0$\times$10$^6$ M$_{\odot}$ \\ 
                & radius         & $r$ & 2.2$\times$10$^{18}$ m \\ 
                & velocity & $v(r)$ & 140 km s$^{-1}$ \\
                & time of circulation & $\dot \Phi^{-1}$     &  4.0$\times$10$^{13}$ s \\ \hline
{\bf potential}$^b$ & {\bf torque}$^2$ & $M$ &{\bf 7.6$\times$10$^{10}$ Nm } \\
                & molecular gas mass   & $m$ & 2.0$\times$10$^6$ M$_{\odot}$ \\ 
                & volume         & $V_o$ & 1.5$\times$10$^{56}$ m$^3$ \\ 
                & total mass     & $m_o = \rho_o V_o$ & 4.3$\times$10$^8$ \\ \hline
{\bf GMC}$^c$   & {\bf torque}$^3$     & $m$ &{\bf 3.2$\times$10$^{44}$ Nm } \\
                & cloud mass     & $m_{GMC}$ & 2.0$\times$10$^6$ M$_{\odot}$ \\ 
                & force          & $F$  & 9.5$\times$10$^{25}$ N \\
                & lever arm      & $l=r$ & 3.3$\times$10$^{18}$ m \\ \hline
\hline 
\end{tabular}
\end{center}
see caption of next table
\end{table}

\begin{table}[htb]
\caption{
Causes for the warp in NGC~1068 continued
}
\begin{center}
\begin{tabular}{lllr}\hline \hline
{\bf gas pressure}$^d$  & {\bf torque }$^4$     & $M$ & {\bf 1.6$\times$10$^{46}$ Nm }   \\
                & particle density &  $\frac{N}{V}$ & 10$^4$ cm$^{-3}$ \\
                & temperature    & $T$ & $1.0 \times 10^5$\,K 		\\ 
                & gas pressure   & $p$ & $1.4 \times 10^{-8}$\, N\,m$^{-2}$ 	\\
                & area           & $A$ & $5.5 \times 10^{35}$\,m$^2$ 		\\ 
                & lever arm      & $l$ & $2.1 \times 10^{18}$\,m 		\\ \hline
{\bf radiation pressure}$^e$ 
	& {\bf torque}$^5$   & $m$ &{\bf $4.8 \times 10^{40}$\,N\,m} 		\\
                & spectral index & $\alpha$ & $\sim -1.0$ 		\\
                & constant & $b$   & $1.0 \times 10^{-17}$\,W\,m$^{-2}$ 	\\
                & luminosity  & $L$    & $6.6 \times 10^{32}$\,W 		\\ 
                & intensity   & $I$   & $1.2 \times 10^{-5}$\,W\,m$^{-2}$ 	\\
            & radiation pressure & $p$ & $3.9 \times 10^{-14}$\,Nm$^{-2}$ 	\\
                & area           & $A$ & $5.5 \times 10^{35}$\,m$^2$ 		\\ 
                & lever arm      & $l$ & $2.1 \times 10^{18}$\,m 		\\ 
\hline \hline
\end{tabular}
\end{center}
Physical reasons for the warp of circum-nuclear gas disks are
summarized in the conclucions (section \ref{eee})
and given in detail in the appendix C in SET99b.
In this table first order estimates are based on:
\\
$^a$ from SET99b, eq.  C3. ;
$^b$ from SET99b, eq.  C2. ;
$^c$ from SET99b, eqs. C1 and C4. ;
$^d$ from SET99b, eqs. C1 and C5. ;
$^e$ from SET99b, eqs. C1 and C6. \\
$^1$ from the parameters of 3DRings for an inner disk, with $r= 108$\,pc \\
$^2$ from the mass of warped nuclear gas disk and a dynamical mass within
     $r<108$\,pc. \\
$^3$ from mass of knot $W$ at $r= 108$\,pc  and its distance 
     relative to the gas disk. \\
$^4$ $N/V$ and $T$ of ionization cone from Capetti et 
     al.\ (1997a) and Gallimore et al.\ (1996a,b) for 
     jet components C and NE.  $A = 0.5'' \times 0.25''$ 
     at $r =1''$, from radio jet map. \\
$^5$ spectral index, constant $b$, and  area ($0.5''
     \times 0.25''$) of jet at $r=72$\,pc from Gallimore et al. (1996b).
\end{table}

\end{document}